\documentclass[twocolumn,showpacs,aps,prd,superscriptaddress]{revtex4-1}

\usepackage{graphicx}
\usepackage{mdwlist}
\usepackage{dcolumn}
\usepackage{multirow}
\usepackage{amsmath}
\usepackage{epsfig}
\usepackage{subfigure}

\newcommand{\BABARPubYear}    {10}
\newcommand{\BABARPubNumber}  {10-030, BAD \#2363, Version 21}

\newcommand{\SLACPubNumber} {14360}

\def\figurebox#1#2#3{%
    \def\arg{#3}%
    \ifx\arg\empty
    {\hfill\vbox{\hsize#2\hrule\hbox to #2{\vrule\hfill\vbox to #1{\hsize#2\vfill}\vrule}\hrule}\hfill}%
    \else
    {\hfill\epsfbox{#3}\hfill}%
    \fi}

\begin{document}

\preprint{\babar-PUB-\BABARPubYear/\BABARPubNumber} 
\preprint{SLAC-PUB-\SLACPubNumber} 

%\begin{flushleft}
%\babar-PUB-\BABARPubYear/\BABARPubNumber\\
%SLAC-PUB-\SLACPubNumber\\
%arXiv:\LANLNumber\ [hep-ex]\\[10mm]
%\end{flushleft}

%%%%%%%%%%%%%%%%%%%%%%%%%%%%%%%%%%%%%%%%%%%%%%%%%%%%%%%%%%%%%%%%%%%%%%%%%%%%%%%%
\title{
{\large \bf
A search for baryon- and lepton-number violating decays of $\Lambda$ hyperons
using the CLAS detector at Jefferson Laboratory
}
}

\newcommand*{\WandJ}{Washington \& Jefferson College, Washington, Pennsylvania 15301}
\newcommand*{\WandJindex}{1}
\affiliation{\WandJ}
\newcommand*{\CMU}{Carnegie Mellon University, Pittsburgh, Pennsylvania 15213}
\newcommand*{\CMUindex}{2}
\affiliation{\CMU}
\newcommand*{\Siena}{Siena College, Loudonville, New York 12220}
\newcommand*{\Sienaindex}{3}
\affiliation{\Siena}

\newcommand*{\ANL}{Argonne National Laboratory, Argonne, Illinois 60439}
\newcommand*{\ANLindex}{4}
\affiliation{\ANL}
\newcommand*{\ASU}{Arizona State University, Tempe, Arizona 85287}%-1504}
\newcommand*{\ASUindex}{5}
\affiliation{\ASU}
\newcommand*{\CSUDH}{California State University, Dominguez Hills, Carson, CA 90747}
\newcommand*{\CSUDHindex}{6}
\affiliation{\CSUDH}
\newcommand*{\CANISIUS}{Canisius College, Buffalo, NY 14208}
\newcommand*{\CANISIUSindex}{7}
\affiliation{\CANISIUS}
\newcommand*{\CUA}{Catholic University of America, Washington, D.C. 20064}
\newcommand*{\CUAindex}{8}
\affiliation{\CUA}
\newcommand*{\CNU}{Christopher Newport University, Newport News, Virginia 23606}
\newcommand*{\CNUindex}{9}
\affiliation{\CNU}
\newcommand*{\SACLAY}{CEA, Centre de Saclay, Irfu/Service de Physique Nucl\'eaire, 91191 Gif-sur-Yvette, France}
\newcommand*{\SACLAYindex}{10}
\affiliation{\SACLAY}
\newcommand*{\UCONN}{University of Connecticut, Storrs, Connecticut 06269}
\newcommand*{\UCONNindex}{11}
\affiliation{\UCONN}
\newcommand*{\FU}{Fairfield University, Fairfield CT 06824}
\newcommand*{\FUindex}{12}
\affiliation{\FU}
\newcommand*{\FIU}{Florida International University, Miami, Florida 33199}
\newcommand*{\FIUindex}{13}
\affiliation{\FIU}
\newcommand*{\FSU}{Florida State University, Tallahassee, Florida 32306}
\newcommand*{\FSUindex}{14}
\affiliation{\FSU}
\newcommand*{\GWUI}{The George Washington University, Washington, DC 20052}
\newcommand*{\GWUIindex}{15}
\affiliation{\GWUI}
\newcommand*{\ISU}{Idaho State University, Pocatello, Idaho 83209}
\newcommand*{\ISUindex}{16}
\affiliation{\ISU}
\newcommand*{\INFNFE}{INFN, Sezione di Ferrara, 44100 Ferrara, Italy}
\newcommand*{\INFNFEindex}{17}
\affiliation{\INFNFE}
\newcommand*{\INFNFR}{INFN, Laboratori Nazionali di Frascati, 00044 Frascati, Italy}
\newcommand*{\INFNFRindex}{18}
\affiliation{\INFNFR}
\newcommand*{\INFNGE}{INFN, Sezione di Genova, 16146 Genova, Italy}
\newcommand*{\INFNGEindex}{19}
\affiliation{\INFNGE}
\newcommand*{\INFNRO}{INFN, Sezione di Roma Tor Vergata, 00133 Rome, Italy}
\newcommand*{\INFNROindex}{20}
\affiliation{\INFNRO}
\newcommand*{\INFNTUR}{INFN, Sezione di Torino, 10125 Torino, Italy}
\newcommand*{\INFNTURindex}{21}
\affiliation{\INFNTUR}
\newcommand*{\ORSAY}{Institut de Physique Nucl\'eaire, CNRS/IN2P3 and Universit\'e Paris Sud, Orsay, France}
\newcommand*{\ORSAYindex}{22}
\affiliation{\ORSAY}
\newcommand*{\ITEP}{Institute of Theoretical and Experimental Physics, Moscow, 117259, Russia}
\newcommand*{\ITEPindex}{23}
\affiliation{\ITEP}
\newcommand*{\JMU}{James Madison University, Harrisonburg, Virginia 22807}
\newcommand*{\JMUindex}{24}
\affiliation{\JMU}
\newcommand*{\KNU}{Kyungpook National University, Daegu 702-701, Republic of Korea}
\newcommand*{\KNUindex}{25}
\affiliation{\KNU}
\newcommand*{\UNH}{University of New Hampshire, Durham, New Hampshire 03824} %-3568}
\newcommand*{\UNHindex}{26}
\affiliation{\UNH}
\newcommand*{\NSU}{Norfolk State University, Norfolk, Virginia 23504}
\newcommand*{\NSUindex}{27}
\affiliation{\NSU}
\newcommand*{\OHIOU}{Ohio University, Athens, Ohio  45701}
\newcommand*{\OHIOUindex}{28}
\affiliation{\OHIOU}
\newcommand*{\ODU}{Old Dominion University, Norfolk, Virginia 23529}
\newcommand*{\ODUindex}{29}
\affiliation{\ODU}
\newcommand*{\RPI}{Rensselaer Polytechnic Institute, Troy, New York 12180-3590}
\newcommand*{\RPIindex}{30}
\affiliation{\RPI}
\newcommand*{\URICH}{University of Richmond, Richmond, Virginia 23173}
\newcommand*{\URICHindex}{31}
\affiliation{\URICH}
\newcommand*{\ROMAII}{Universita' di Roma Tor Vergata, 00133 Rome Italy}
\newcommand*{\ROMAIIindex}{32}
\affiliation{\ROMAII}
\newcommand*{\MSU}{Skobeltsyn Institute of Nuclear Physics, Lomonosov Moscow State University, 119234 Moscow, Russia}
\newcommand*{\MSUindex}{33}
\affiliation{\MSU}
\newcommand*{\SCAROLINA}{University of South Carolina, Columbia, South Carolina 29208}
\newcommand*{\SCAROLINAindex}{34}
\affiliation{\SCAROLINA}
\newcommand*{\TEMPLE}{Temple University,  Philadelphia, PA 19122 }
\newcommand*{\TEMPLEindex}{35}
\affiliation{\TEMPLE}
\newcommand*{\JLAB}{Thomas Jefferson National Accelerator Facility, Newport News, Virginia 23606}
\newcommand*{\JLABindex}{36}
\affiliation{\JLAB}
\newcommand*{\UTFSM}{Universidad T\'{e}cnica Federico Santa Mar\'{i}a, Casilla 110-V Valpara\'{i}so, Chile}
\newcommand*{\UTFSMindex}{37}
\affiliation{\UTFSM}
\newcommand*{\EDINBURGH}{Edinburgh University, Edinburgh EH9 3JZ, United Kingdom}
\newcommand*{\EDINBURGHindex}{38}
\affiliation{\EDINBURGH}
\newcommand*{\GLASGOW}{University of Glasgow, Glasgow G12 8QQ, United Kingdom}
\newcommand*{\GLASGOWindex}{39}
\affiliation{\GLASGOW}
\newcommand*{\VT}{Virginia Tech, Blacksburg, Virginia 24061} %-0435}
\newcommand*{\VTindex}{40}
\affiliation{\VT}
\newcommand*{\VIRGINIA}{University of Virginia, Charlottesville, Virginia 22901}
\newcommand*{\VIRGINIAindex}{41}
\affiliation{\VIRGINIA}
\newcommand*{\WM}{College of William and Mary, Williamsburg, Virginia 23187} %-8795}
\newcommand*{\WMindex}{42}
\affiliation{\WM}
\newcommand*{\YEREVAN}{Yerevan Physics Institute, 375036 Yerevan, Armenia}
\newcommand*{\YEREVANindex}{43}
\affiliation{\YEREVAN}

\newcommand*{\NOWJLAB}{Thomas Jefferson National Accelerator Facility, Newport News, Virginia 23606}
\newcommand*{\NOWODU}{Old Dominion University, Norfolk, Virginia 23529}
\newcommand*{\NOWEDINBURGH}{Edinburgh University, Edinburgh EH9 3JZ, United Kingdom}
\newcommand*{\NOWASU}{Arizona State University, Tempe, Arizona 85287}
\newcommand*{\EMAIL}{mmccracken@washjeff.edu}
 %%%%%%%%%%%%%%% END OF Latex Macros for institute addresses  %%%%%%%%%%%%%%%%%%%%%%%%% 

\author{M.E.~McCracken}
\altaffiliation[E-mail: ]{\EMAIL}
\affiliation{\WandJ}
\affiliation{\CMU}
\author{M.~Bellis}
\affiliation{\Siena}
\author {K.P.~Adhikari} 
\affiliation{\ODU}
\author {D.~Adikaram} 
\altaffiliation[Current address: ]{\NOWJLAB}
\affiliation{\ODU}
\author {Z.~Akbar} 
\affiliation{\FSU}
\author {S.~Anefalos~Pereira} 
\affiliation{\INFNFR}
\author {R.A.~Badui} 
\affiliation{\FIU}
\author {J.~Ball} 
\affiliation{\SACLAY}
\author {N.A.~Baltzell} 
\altaffiliation[Current address: ]{\NOWJLAB}
\affiliation{\ANL}
\affiliation{\SCAROLINA}
\author {M.~Battaglieri} 
\affiliation{\INFNGE}
% ...

\collaboration{The CLAS Collaboration}
\noaffiliation

\date{\today}% It is always \today, today, but you may specify any date with \date.
%%%%%%%%%%%%%%%%%%%%%%%%%%%%%%%%%%%%%%%%%%%%%%%%%%%%%%%%%%%%%%%%%%%%%%%%%%%%%%%%
%%%%%%%%%%%%%%%%%%%%%%%%%%%%%%%%%%%%%%%%%%%%%%%%%%%%%%%%%%%%%%%%%%%%%%%%%%%%%%%%
\begin{abstract}
We present a search for ten baryon-number violating decay modes of $\Lambda$ hyperons using the CLAS detector at Jefferson Laboratory. 
%Nine of these decay modes result in a single meson and single lepton in the final state ($\Lambda \rightarrow m \ell$), reactions that violate both baryon number ($B$) and lepton number ($L$), but conserve either the sum or the difference of $B$ and $L$ ({\em i.e.}, $B\pm L$). 
Nine of these decay modes result in a single meson and single lepton in the final state ($\Lambda \rightarrow m \ell$) and conserve either the sum or the difference of baryon and lepton number ($B \pm L$).
The tenth decay mode ($\Lambda \rightarrow \bar{p}\pi^+$) represents a difference in baryon number of two units and no difference in lepton number.
We observe no significant signal and set upper limits on the branching fractions of these reactions in the range $(4-200)\times 10^{-7}$ at the 90\% confidence level.
\end{abstract}
%%%%%%%%%%%%%%%%%%%%%%%%%%%%%%%%%%%%%%%%%%%%%%%%%%%%%%%%%%%%%%%%%%%%%%%%%%%%%%%%
%%%%%%%%%%%%%%%%%%%%%%%%%%%%%%%%%%%%%%%%%%%%%%%%%%%%%%%%%%%%%%%%%%%%%%%%%%%%%%%%
% PACS, the Physics and Astronomy Classification Scheme.
\maketitle
%%%%%%%%%%%%%%%%%%%%%%%%%%%%%%%%%%%%%%%%%%%%%%%%%%%%%%%%%%%%%%%%%%%%%%%%%%%%%%%%
%%%%%%%%%%%%%%%%%%%%%%%%%%%%%%%%%%%%%%%%%%%%%%%%%%%%%%%%%%%%%%%%%%%%%%%%%%%%%%%%
% The body of the paper starts here
%%%%%%%%%%%%%%%%%%%%%%%%%%%%%%%%%%%%%%%%%%%%%%%%%%%%%%%%%%%%%%%%%%%%%%%%%%%%%%%%
\section{Introduction} \label{sec:introduction}
%%%%%%%%%%%%%%%%%%%%%%%%%%%%%%%%%%%%%%%%%%%%%%%%%%%%%%%%%%%%%%%%%%%%%%%%%%%%%%%%
% Motivation
%%%%%%%%%%%%%%%%%%%%%%%%%%%%%%%%%%%%%%%%%%%%%%%%%%%%%%%%%%%%%%%%%%%%%%%%%%%%%%%%

The Standard Model (SM) of particle physics~\cite{Glashow:1961tr,Weinberg:1967tq,Salam:1968rm}
has had great success in interpreting and predicting experimental results since its conception in the late 1960s.
There are, however, features of our universe that are inconsistent with the SM framework. 
Astronomical observations suggest that our universe is dominated by matter over antimatter~\cite{Coppi:2004za, Steigman:1976ev}.
Sakharov proposed in 1967 that this asymmetry suggests fundamental interactions that violate CP-symmetry and baryon-number conservation~\cite{Sakharov:1967dj}.
The observed quark-sector $CP$ violation, combined with baryon-number violating (BNV) processes that {\em are} allowed by the Standard Model, are insufficient~\cite{Kuzmin:1985mm} to account for the observed matter-antimatter asymmetry in our universe. 
A possible explanation for this discrepancy is that there are yet-unobserved interactions that violate baryon-number conservation.

Baryon-number violating reactions are features of several theoretical extensions to the Standard Model, perhaps most notably the $SU(5)$ Grand Unified Theory (GUT) of Georgi and Glashow~\cite{Georgi:1974sy}, $SU(5)$ being the larger gauge group in which the Standard Model's $SU(3)\times SU(2)$ are embedded~\cite{chengli1984}.
The $SU(5)$ theory proposes the existence of two new gauge bosons, the $X$ and $Y$ leptoquarks, so called because they allow vertices such as $q\rightarrow X\ell$, where $q$ is a quark and $\ell$ a lepton.
Other experiments have been performed to search for BNV processes in decays of the nucleon~\cite{Nishino:2009aa, Abe:2014, PDG-2014},
$\tau$ leptons~\cite{Godang:1999ge,Miyazaki:2005ng,Aaij:2013fia},
top quarks~\cite{Chatrchyan:2013bba},
hadrons with bottom and charm quarks~\cite{BABAR:2011ac,Rubin:2009aa}, and the $Z$ boson~\cite{Abbiendi:1998nj}, but no signal has yet been observed. 
The most stringent limits on such processes come from nucleon decays, and these have been used to constrain BNV decays in higher-generation ({\em i.e.}, $c$, $b$, and $t$) quarks~\cite{Hou:2005iu}.
However, multiple amplitudes can contribute to a given decay process and these amplitudes can interfere either constructively
or destructively, depending on their relative phases. The theoretical calculations that constrain these BNV processes do not take into account interference between the
amplitudes due to the large parameter space. This allows for non-observation in one mode ({\em e.g.},
decays involving the $u$ or $d$ quark) while still being consistent with observable BNV processes in some other mode ({\em e.g.}, coupling to another quark flavor).

Here we present a search for baryon-number ($B$) and lepton-number ($L$) non-conserving decays of the $\Lambda$ hyperon as a direct probe of couplings of BNV interactions to the strange quark.
%Combining the branching fractions and uncertainties of six observed decay modes of the $\Lambda$ listed by the PDG suggests that the {\em total} branching fraction of these falls in the range
By summing the branching fractions and experimental uncertainties (in quadrature) of the six observed $\Lambda$ decay modes~\cite{PDG-2014}, we find the total branching fraction to be $1.001 \pm 0.007$, implying that there is room for yet-unobserved decay modes.

We investigated eight decay modes in which the $\Lambda$ decays to a charged meson and a charged lepton, conserving charge in all decays. The meson is either a $\pi^{\pm}$ or $K^\pm$ and the lepton is either a $e^{\mp}$ or a $\mu^{\mp}$. 
We produced the $\Lambda$ by means of a photon beam incident on a liquid hydrogen target through the exclusive reaction $\gamma p\rightarrow K^+ \Lambda$. 
For these eight modes we can completely reconstruct the three final state particles. 
We also searched for the decay of a $\Lambda$ to a $K^{0}_{S}$ and a neutrino, which must be inferred from the missing momentum. 
%All nine of these modes violate both baryon number and lepton number. 
Selection of these nine channels is motivated by searching for decay of the $\Lambda$ to a lighter pseudo-scalar meson.
In each case the final-state meson is included to preserve charge and angular momentum conservation.
Thus, violation of $L$ in these reactions is a consequence of $B$ violation rather than a primary motivation.

In addition, we searched for the BNV
decay of the $\Lambda$ to an anti-proton and a $\pi^+$, for which we can completely reconstruct the final state. 
The $\Lambda \rightarrow \bar{p}\pi^+$ reaction presents an opportunity to search for $\Lambda$-$\bar{\Lambda}$ oscillations, {\em i.e.} a process by which the $\Lambda$ oscillates into its anti-particle counterpart ($\bar{\Lambda}$), which then undergoes a standard-model decay ($\bar{\Lambda}\rightarrow \bar{p}\pi^+$).
Though there is not a simple way to picture this reaction proceeding via $X$ boson coupling, theoretical and experimental ({\em e.g.}~\cite{Serebrov2008181,Abe:2011ky}) work has been performed by other groups looking for similar oscillations of the neutron.  
Such baryon-antibaryon oscillations would have far-reaching implications and are often held up as evidence for high-energy theories ranging from see-saw models~\cite{babu,dutta} to extra dimensions~\cite{nussinov}.

The properties of these reactions are summarized in Table~\ref{t:reactions}.  
These specific decays were chosen for several reasons: 
\begin{itemize*}
\item Each reaction shows evidence of $\Delta B \neq 0$ and/or $\Delta L=\pm 1$.
\item Each reaction conserves electric charge and angular momentum.
\item This selection includes reactions that either preserve or violate $B-L$, a conserved quantity proposed by several GUTs \cite{WilczekZee}.
\item The CLAS detector is optimized to reconstruct the final-state charged particles produced in each reaction, except for the neutrino
    which can be inferred by calculating the missing 4-momentum in the event.
\end{itemize*}

\begin{table}
\begin{center}
  \begin{tabular}{|l|c|c|c|c|}\hline
  Decay & $\Delta B$ & $\Delta L$ & $\Delta(B-L)$ & detected \\ \hline 
    $ \Lambda \rightarrow K^+ \ell^-$ & $-1$ & $+1$ & $-2$ & $K^{+}_{r},K^{+}_{d},\ell^-$ \\
    $ \Lambda \rightarrow K^- \ell^+$ & $-1$ & $-1$ & $0$ & $K^{+}_{r},K^{-}_{d},\ell^+$ \\
    $ \Lambda \rightarrow \pi^+ \ell^-$ & $-1$ & $+1$ & $-2$ & $K^{+}_{r},\pi^{+},\ell^-$ \\
    $ \Lambda \rightarrow \pi^- \ell^+$ & $-1$ & $-1$ & $0$ & $K^{+}_{r},\pi^{-},\ell^+$ \\
    $ \Lambda \rightarrow K^{0}_{S} \nu$, $K^{0}_{S}\bar{\nu}$ & $-1$ & $\pm1$ & $0$, $-2$ & $K^{+}_{r},\pi^{+},\pi^-$\\
    $ \Lambda \rightarrow \bar{p}\pi^+$ & $-2$ & $0$ & $-2$ & $K^{+}_{r},\bar{p},\pi^+$ \\ \hline
\end{tabular}
\end{center}
\caption[Properties of conserved quantities and couplings for each BNV channel]{Properties of conserved quantities for each $\Lambda$ BNV channel.  Here $\ell$ represents a lepton, either $e$ or $\mu$.  The rightmost column shows the detected final-state particles for each channel.  Subscripts denote recoil ($r$) or decay ($d$) kaons.}
\label{t:reactions}
\end{table}

%%%%%%%%%%%%%%%%%%%%%%%%%%%%%%%%%%%%%%%%%%%%%%%%%%%%%%%%%%%%%%%%%%%%%%%%%%%%%%%%

\subsection*{Analysis overview\label{sec:analysis}}
%%%%%%%%%%%%%%%%%%%%%%%%%%%%%%%%%%%%%%%%%%%%%%%%%%%%%%%%%%%%%%%%%%%%%%%%%%%%%%%%
% Overview of analysis
%%%%%%%%%%%%%%%%%%%%%%%%%%%%%%%%%%%%%%%%%%%%%%%%%%%%%%%%%%%%%%%%%%%%%%%%%%%%%%%%
This analysis was performed in three stages.  Here, we present a brief outline; details of each stage will be given in the following sections.

%\begin{enumerate}
%\item 
{\boldmath $\Lambda$} {\bf identification}.  In order to assess the sensitivity of our study, we first determined the number of $\Lambda$ hyperons produced during the run period. 
We did so by considering the charged decay mode, $\Lambda \rightarrow p \pi^-$.
%We begin with the entire \textit{g11} dataset, which contains roughly 20 billion triggers.  
%Events corresponding to the photoproduction of a $\Lambda$ represent only a small fraction of these ($\approx0.01$\%).  
We applied a set of simple cuts on kinematic observables and timing for the recoil $K^{+}$ to effectively identify potential $\Lambda$ events.
%This stage of the analysis is common to all of the nine reactions under study.  
%Assuming that the bnv contribution to these events is small, the vast majority of these events correspond to $\Lambda\rightarrow p \pi^-$.  
We used this sample to determine the total number of $\gamma p \rightarrow K^+\Lambda \rightarrow K^+ p \pi^-$ events detected, and then acceptance corrected to find the total number of $\gamma p \rightarrow K^+\Lambda\rightarrow K^+ p \pi^-$ events produced during the run period. 

%\item 
{\bf Channel-specific tuning}.  When searching for evidence of the BNV decays listed in Table~\ref{t:reactions}, we performed a blind search. 
We developed a set of background separation cuts for each BNV channel, based on timing information for all charged final-state particles and kinematic observables for the event.
%In developing these cuts, we balance the optimization of kinematic cuts for both discovery and upper-limit sensitivity through use of the Punzi figure of merit \cite{punzi}, which requires investigation of cut effects on signal and background.  
In developing these cuts, we balanced the optimization of kinematic cuts for both discovery and upper-limit sensitivity by maximizing a figure of merit (approximately the BNV signal efficiency divided by the square root of the number of background events).
%To avoid bias, we optimize cut efficiency using Monte Carlo BNV signal events while blinding the signal regions of data plots.
We assessed the signal efficiency using a Monte Carlo technique and the background size using side-bands of the blinded signal region.
This step provided a set of cuts for each BNV channel that reduces background and provides optimal analysis power.
%At this point, we can investigate the sensitivity of the analysis (\textit{i.e.}, the smallest bnv signal that would be identifiable).

%\item 
{\bf Unblinding}.  We then unblinded the signal regions of kinematic variable plots, and determined whether a signal is present.  For nine of the decays, the expected backgrounds are 0 or 1 event and so we used the Feldman-Cousins method \cite{feldcous} to determine upper bounds on branching fractions; for the remaining channel, the backgrounds are higher and so we scanned the relevant parameters in our fit to determine the 90\% coverage for the number of signal events
in the dataset.
%\end{enumerate}

\section{\boldmath The CLAS detector and dataset\label{sec:clas}}
%%%%%%%%%%%%%%%%%%%%%%%%%%%%%%%%%%%%%%%%%%%%%%%%%%%%%%%%%%%%%%%%%%%%%%%%%%%%%%%%
% Dataset and detector
%%%%%%%%%%%%%%%%%%%%%%%%%%%%%%%%%%%%%%%%%%%%%%%%%%%%%%%%%%%%%%%%%%%%%%%%%%%%%%%%

The CLAS detector is described in detail elsewhere~\cite{Mecking:2003zu}. 
The dataset comes from the {\it g11} run period, which collected data during May and June, 2004. 
A bremsstrahlung photon beam was produced by a 4.023-GeV electron beam incident on a gold radiator. 
Electrons were provided by CEBAF (Continuous Electron Beam Accelerator Facility) in 2-ns bunches. 
Photon energy and timing information were provided by a tagging spectrometer which directs the electrons after the radiator through a magnetic field and onto a set of scintillators, providing photon energy resolution of 4.0~MeV.

For this analysis we made use of the CLAS drift chamber and toroidal magnet systems to measure the momenta of the charged final-state particles.
Velocity measurements are made by a start counter (consisting of scintillators placed within 11.6~cm of the target) and a set of time-of-flight (TOF) scintillators located approximately 4~m from the target. 
Timing information from the photon tagger is combined with this system to calculate a velocity %$\beta$, 
for each charged track and this is compared with the particle hypothesis in the particle identification (PID) algorithm. 
More details of the g11 dataset, calibration procedures, and systematic studies can be found in ({\em e.g.})~\cite{mccracken}.

%\section{\boldmath Overview of analysis\label{sec:analysis}}
%\input{overview_of_analysis.tex}

\section{\boldmath Identification of $\gamma p \rightarrow K^+\Lambda \rightarrow K^+p\pi^-$ events\label{sec:stdmod}}
%%%%%%%%%%%%%%%%%%%%%%%%%%%%%%%%%%%%%%%%%%%%%%%%%%%%%%%%%%%%%%%%%%%%%%%%%%%%%%%%
% Standard model identification
%%%%%%%%%%%%%%%%%%%%%%%%%%%%%%%%%%%%%%%%%%%%%%%%%%%%%%%%%%%%%%%%%%%%%%%%%%%%%%%%
In order to compare the branching fraction for a BNV decay mode to that of the standard-model $\Lambda\rightarrow p \pi^-$ decay, we must first assess the number of standard-model decays that occurred during the data-taking period.
We did so by investigating the (exclusive) $\gamma p \rightarrow K^+ \Lambda\rightarrow K^+ p \pi^-$ reaction.
Earlier studies \cite{mccracken} have shown that the $\gamma p \rightarrow K^+ \Lambda$ signal is easily separable from background when all three final-state particles are reconstructed.

Because a different set of background separation cuts will be applied to the BNV channels, we must correct the number of reconstructed signal events ($N_{\textrm{rec}}$) to find $N_{\textrm{prod}}$, the number of $\gamma p \rightarrow K^+\Lambda \rightarrow K^+ p \pi^-$ events that were produced in CLAS during the run period.
In order to calculate the efficiency of the detector and analysis cuts, we generated $3\times 10^6$ Monte Carlo (MC) $\gamma p \rightarrow K^+ \Lambda$ events, and weighted the distribution of these events in $\cos\theta_{cm}^{K}$ (the $K^+$ production angle in the center-of-mass frame) according to published $d\sigma/d\cos\theta$ data~\cite{mccracken}.
Photon energies between 0.909~GeV (threshold for $K^+\Lambda$ production) and 3.860~GeV (the upper-limit of the photon tagger during data taking) were generated from a bremsstrahlung spectrum given incident electrons of energy 4.023~GeV (matching the \textit{g11} run conditions).
%We then used the GEANT software suite to model the $\Lambda \rightarrow p \pi^-$ decay, according only to phase space constraints, and to model the acceptance of CLAS.
We then used the collaboration-standard GEANT-based~\cite{Brun:1994aa} software suite (GSIM) to model the CLAS acceptance, allowing GEANT to produce the $\Lambda \rightarrow p \pi^-$ decay according only to phase space constraints.

We began the data reduction by selecting from the dataset all events in which three reconstructed final-state tracks, two of positive charge and one of negative charge, were coincident with a tagged photon.
To each event, we assign the mass hypothesis consistent with $K^+$, $p$, and $\pi^-$ final-state particles, selecting the permutation of positive tracks as that with the value of the invariant mass of the $p$ and $\pi^-$ candidate tracks, $INV(p,\pi^-)$, closest to the nominal $\Lambda$ mass.
We then apply a loose cut on the square of the total missing mass, $MM^2$ of each event, keeping events for which $MM^2 \in [-0.002, 0.0005]$~GeV$^{2}/c^4$.

\begin{figure}
\begin{center}
\includegraphics[width=0.5\textwidth]{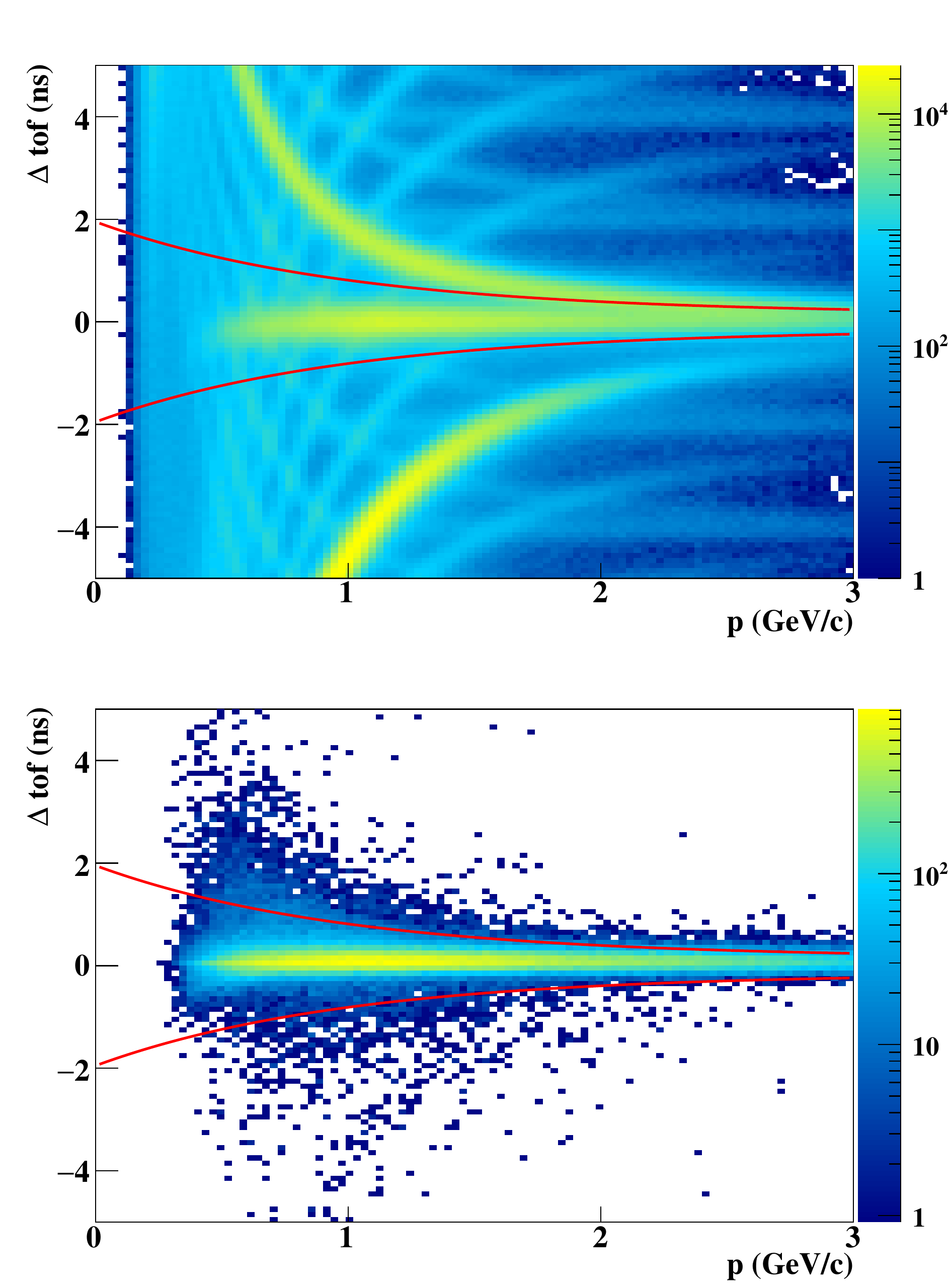}
\caption[$\Delta tof$ versus $|\vec{p}|$ for all positive MC tracks]{(Color Online) $\Delta tof$ versus momentum (p) for $K^+$ candidate tracks for data (top) and MC (bottom) that pass the cut on $MM^2$.
The red curves show the upper and lower bounds of the $K^+$ PID cuts.
Several additional structures are apparent in data distribution.
The band of events with $\Delta tof \gtrsim 1$~ns represents $\mu^+$ from $K^+$ decay, that with $\Delta tof \lesssim 1$~ns represents protons, and
the less-populated bands represent events matched with photons from a different 2-ns beam bunch.
The majority of these are removed by further cuts.}
\label{f:dtof}
\end{center}
\end{figure}

To the hypothetical $K^+$ track we applied a PID cut based on timing and momentum information from CLAS.
CLAS measures the time-of-flight, $tof_{m}$, {\em i.e.} the time elapsed between the primary event vertex and the track's triggering of the TOF scintillators.
We also calculate a hypothetical time-of-flight, $tof_{h}$, based on our mass hypothesis and tracking information:
\begin{equation}
tof_{h} = \frac{d}{\beta c} = \frac{d}{c}\left[1+\frac{m_{h}^2c^2}{p^2} \right]^{\frac{1}{2}},
\end{equation}
where $p$ is the track's momentum (determined by its curvature through the magnetic field of CLAS); $\beta$ is related to the track's velocity (determined from tracking and timing information), $\beta=v/c$; $m_{h}$ is the hypothetical mass for the track ($493.7$~MeV/$c^2$ for the $K^+$); $d$ is the path length of the track from the vertex to the TOF system, and $c$ is the speed of light. 
This PID cut (and later cuts) is based on the difference of these two quantities,
\begin{equation}\label{e:dtof}
\Delta tof = tof_{h} - tof_{m}.
\end{equation}
In the case where the mass hypothesis is correct, we expect the measured and hypothetical $tof$ to be roughly identical (modulo timing resolution), and thus $\Delta tof\approx 0$.
If the hypothetical mass is greater (less) than the particle's actual mass, then we expect the $\Delta tof$ to be greater (less) than zero.
We found that suitable separation of $K^+$ candidate tracks from non-$K^+$ tracks is achieved by a two-dimensional cut, keeping events for which
\begin{equation}
|\Delta tof| \leq (1.8\textrm{ ns})\exp \left(-\frac{p}{1\textrm{ GeV}/c}\right) + 0.15\textrm{ ns}.
\end{equation}
The $\Delta tof$ versus $|\vec{p}|$ plane for signal Monte Carlo events is shown in Fig.~\ref{f:dtof}.

\begin{figure*}
\begin{center}
\includegraphics[width=0.98\textwidth]{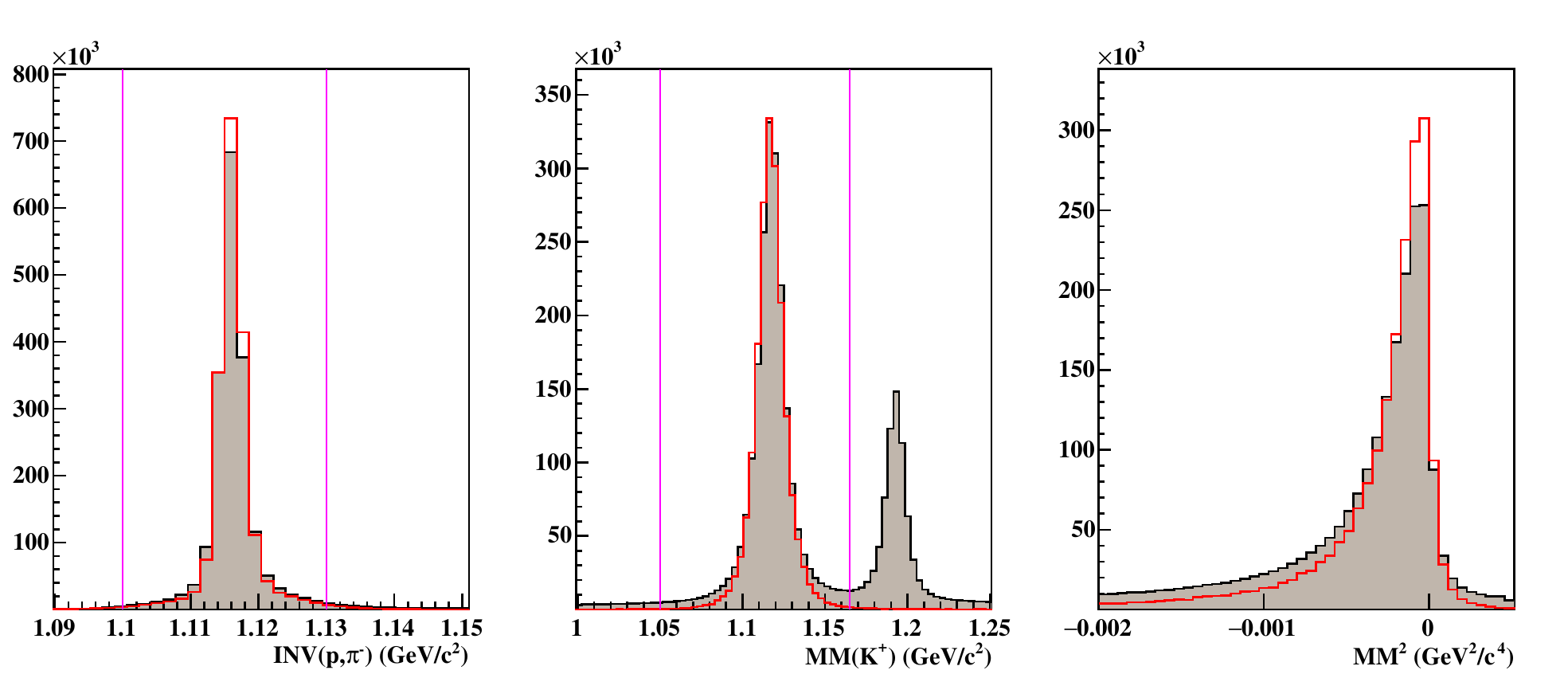}
\caption[Comparison of std mod MC and data]{(Color Online) $INV(p,\pi^-)$ (left), $MM(K^+)$ (center), and $MM^2$ (right) distributions for standard-model data (filled) and MC events (unfilled).
In the $INV(p,\pi^-)$ histograms the distributions are of events that pass all cuts, and the vertical lines show the limits used in signal counting.
In the $MM(K^+)$ histograms, the distributions are of events that pass all but the cut on $MM(K^+)$ (thus, the $\Sigma^0$ peak is still visible in data); the vertical lines show the limits of the cut on $MM(K^+)$.
In the $MM^2$ histograms, the distributions are of events that pass all cuts.}
\label{f:mc_data_compare}
\end{center}
\end{figure*}

We then make a further cut on the missing mass off of the $K^+$,
\begin{equation}
MM(K^+) \equiv (p_{\gamma} + p_{\textrm{t}} - p_{K})^{2},
\end{equation}
where $p_{\gamma}$, $p_{\textrm{t}}$, and $p_{K}$ are the four-momenta of the incident photon, target proton, and recoil $K^+$, respectively.
We kept events for which $MM(K^+)$ is in the range $[1.05\textrm{ GeV},1.165\textrm{ GeV}]$.  
In addition to non-strange backgrounds, this cut removes contamination from photoproduction of higher-mass hyperons that include $\Lambda$ in their decay chain (predominantly $\gamma p \rightarrow K^+ \Sigma^0 \rightarrow K^+ \gamma \Lambda$).
We also applied geometrical fiducial cuts, omitting events from regions of the detector for which our simulation is inaccurate.
After all cuts, we found that data and MC distributions are quite similar, as shown in Fig.~\ref{f:mc_data_compare}.

\begin{figure}
\begin{center}
\includegraphics[width=0.48\textwidth]{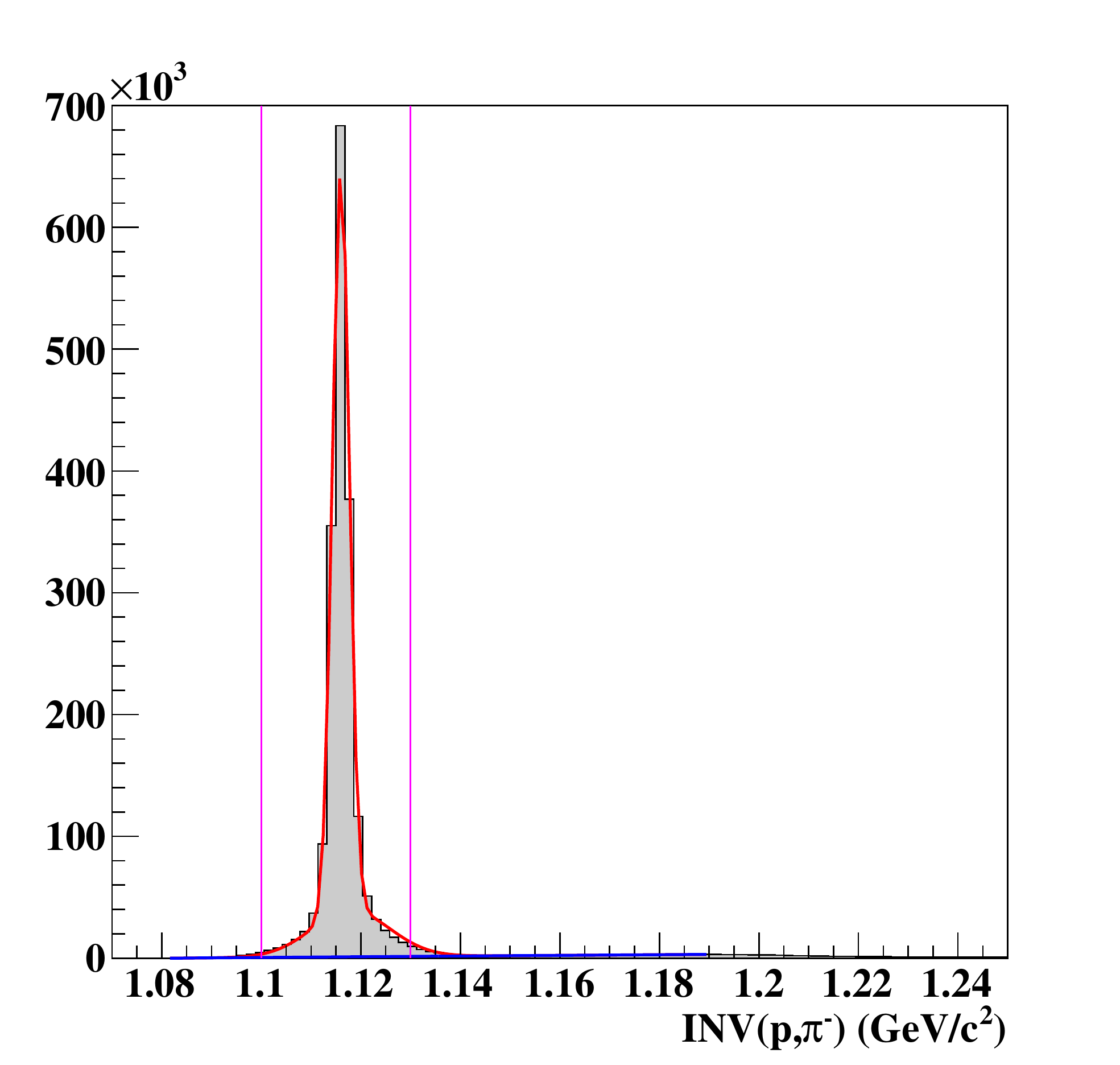}\\
\includegraphics[width=0.48\textwidth]{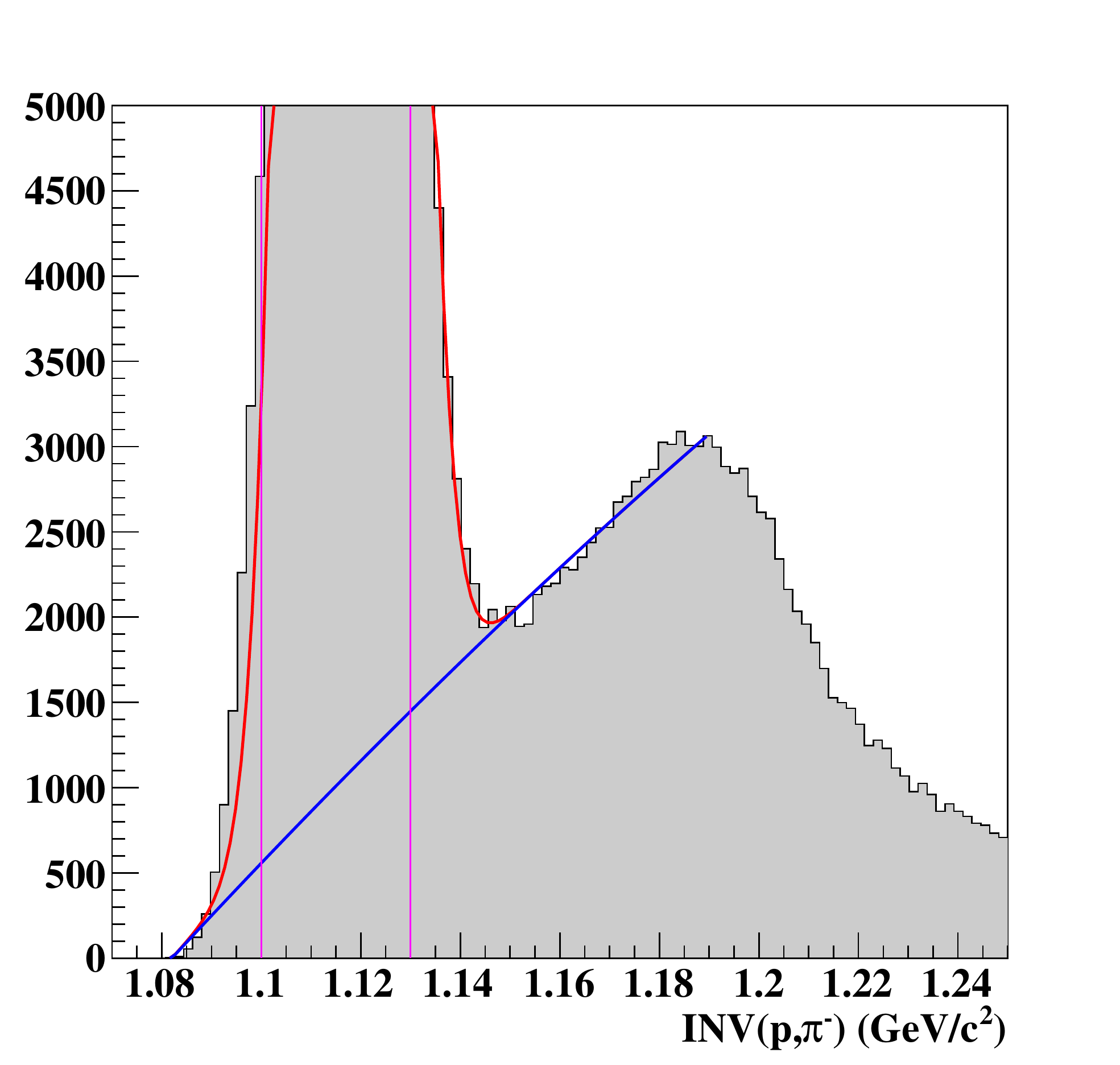}
\caption[Fit to $INV(p,\pi^-)$ for all events passing $\Lambda$ ID cuts]{(Color Online) Fit to the $INV(p,\pi^-)$ distributions for all events passing the standard model decay identification cuts. The lower plot shows the same distribution as the upper, but with limited vertical scale.  The red curve shows the full (signal and background) fit function; the blue curve shows only the polynomial background.  The magenta lines show the limits of integration for the event count.}
\label{f:invmstdmod}
\end{center}
\end{figure}

After these cuts, we identified the signal events by inspecting a histogram of $INV(p,\pi^-)$.  
Fig.~\ref{f:invmstdmod} shows that this distribution is exceptionally clean.
We modeled these data with a third-order-polynomial for background processes and a double-Gaussian for signal processes.
We then extracted the number of signal events by taking the excess of the data histogram above the background function in all of the histogram bins within 0.015~GeV/$c^2$ of the nominal $\Lambda$ mass, yielding $N_{\textrm{rec}} = 1.861\times10^6$ reconstructed $\Lambda\rightarrow p \pi^-$ signal events.
Because the shape and magnitude of the background and the magnitude of the signal are dependent on kinematics, we vetted the above estimate by separating the data into ten bins in $\cos\theta_{cm}^{K}$ and performing the fit in each bin.  This method again yields an estimate of $N_{\textrm{rec}} = 1.861\times10^6$.

\subsection*{Acceptance correction}

With the number of reconstructed $\gamma p \rightarrow K^+\Lambda \rightarrow K^+ p \pi^-$ events in hand, we could then correct for the effects of the detector's acceptance and the efficiency of our analysis cuts to estimate $N_{\textrm{prod}}$.
We began by applying the cuts described above to the standard-model Monte Carlo (MC).
Because the generation of MC events matches the data in photon energy distribution and kinematics, we separated the data and MC coarsely into ten bins in $\cos\theta_{cm}^{K}$.
In each bin, we calculated the acceptance $\epsilon_{p\pi^-}(\cos\theta_{cm}^{K})$ by simply dividing the number of MC events that pass CLAS simulation and acceptance cuts by the total number of MC events generated.
We used this factor to determine the number of $K^+\Lambda\rightarrow K^+ p \pi^-$ events produced during the run in each $\cos\theta_{cm}^{K}$ bin, and completed the calculation by summing over the angular bins:
\begin{eqnarray}
N_{\textrm{prod}} &=& \displaystyle\sum_{\cos\theta_{cm}^{K}}\frac{N_\textrm{rec}(\cos\theta_{cm}^{K})}{\epsilon_{p\pi^-}(\cos\theta_{cm}^{K})} \\
           &=& 3.71\times10^7\label{e:nprod}
\end{eqnarray}
%\begin{equation}
%\frac{3.006\times10^6}{\eta_{p\pi^-}} = 6.227\times10^7.
%\end{equation}

\section{\boldmath BNV candidate selection and optimization\label{sec:selection}}
%%%%%%%%%%%%%%%%%%%%%%%%%%%%%%%%%%%%%%%%%%%%%%%%%%%%%%%%%%%%%%%%%%%%%%%%%%%%%%%%
% Candidate selection and optimization
%%%%%%%%%%%%%%%%%%%%%%%%%%%%%%%%%%%%%%%%%%%%%%%%%%%%%%%%%%%%%%%%%%%%%%%%%%%%%%%%
Because of the sensitive nature of the possibility of BNV discovery, we pursued a \textit{blind analysis}.  
For each of the BNV $\Lambda \rightarrow AB$ decay modes under investigation (see Table \ref{t:reactions}), we tuned a set of cuts, $C$, using
the kinematic quantities $MM(K^+)$, $INV(A,B)$, and $MM^2$, as well as the timing information for $A$ and $B$. 
As before with the Standard Model $\Lambda$ decay, we identified BNV signal using the $INV(A,B)$ spectrum (except for the $\Lambda \rightarrow K^{0}_{s}\nu$ channel described below). 
In tuning each set of cuts, we tried to strike a balance between reducing the large number of non-signal events and maintaining acceptance for a potentially small BNV signal. 

Punzi~\cite{punzi} has proposed a figure of merit for performing such optimizations, which has been used in several other searches for rare reactions. For a set of cuts $C$ the Punzi figure of merit, $\mathcal{P}(C)$, is defined as
\begin{equation}
\mathcal{P}(C) = \frac{\epsilon(C)}{b^2 + 2 a \sqrt{B(C)} + b\sqrt{b^2 + 4a\sqrt{B(C)} + 4B(C)}},
\end{equation}
where $\epsilon(C)$ is the efficiency of the cuts when applied to signal, $B(C)$ is the number of background events passing cuts $C$, and $a$ and $b$ are the number of standard deviations corresponding to the analysis-defined significance and statistical power.
In this analysis, we chose $b = a = 4$, indicating a $4\sigma$ confidence level.
With these choices, the figure of merit simplifies to 
\begin{equation}\label{eq:punzifom}
\mathcal{P}(C) = \frac{\epsilon(C)}{16\left( 2 + \sqrt{B(C)}\right)}.
\end{equation}

We have tuned our cuts by simultaneously assessing $\epsilon(C)$ using MC BNV events and $B(C)$ from side-bands of the blinded signal region.
In all plots that would identify any BNV signal (\textit{e.g.}, $INV(A,B)$) we blind the signal region.
We postpone the unblinding of the signal regions of all of our data plots until after the optimization of the analysis cuts, once we are confident that all cuts and systematic effects are understood.

For each BNV $\Lambda \rightarrow AB$ decay under investigation, we generate $10^6$ $\gamma p \rightarrow K^+\Lambda \rightarrow K^+AB$ Monte Carlo events, matching photon energies to the run conditions and $K^{+}\Lambda$ kinematics to the measured $d\sigma/d\cos\theta$ (as for the standard model MC).
We generate kinematics for the $\Lambda\rightarrow AB$ decays according only to phase-space constraints.  
In the case of the $\Lambda \rightarrow K^0_s \nu$ reaction, the subsequent $K^0_s\rightarrow \pi\pi$ decay is modeled by GEANT at the time of detector simulation.  
%We then run these MC events through the full CLAS detector simulation.

For each channel, we choose the positive track mass hypothesis that yields a value of $INV(A,B)$ nearest to the nominal $\Lambda$ mass (as was done for the standard-model analysis above).
The analysis cuts for each channel begin with a PID cut on $\Delta tof$ (see eq.~\ref{e:dtof}) for each final-state particle.
For each particle type, we apply a loose two-sided cut in the $\Delta tof$ \textit{vs} $p$ plane, similar to that applied to the $K^+$ track for the SM decay.  
CLAS resolution allows us to make these cuts loose; however, the 
characteristic decay length of $K^\pm$ is on the same order as the dimensions of CLAS and the non-trivial fraction of $K^\pm$ that decay in the detector results in these cuts reducing the data sample by approximately half (depending on the number of charged kaons detected for each channel).

\begin{figure*}
\begin{center}
\includegraphics[width=1.0\textwidth]{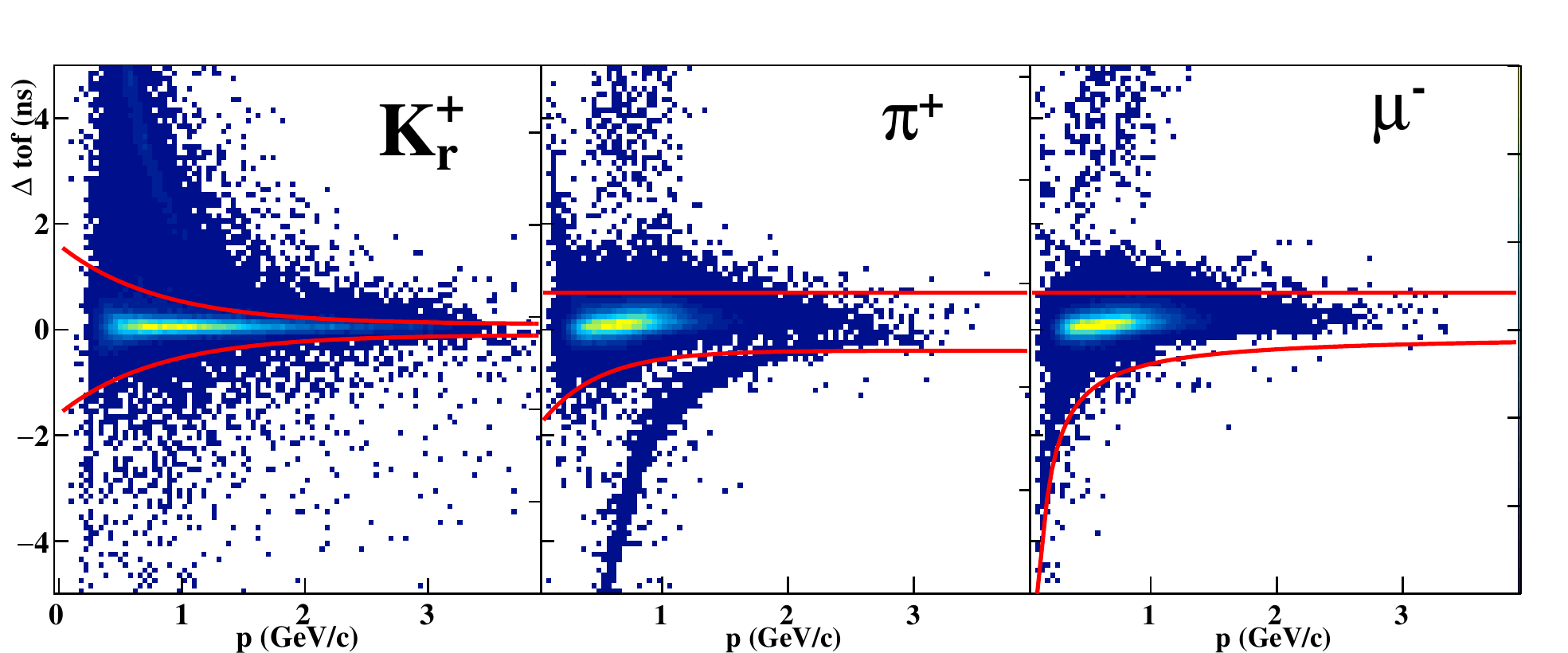}
\caption[$\Delta tof$ versus $|\vec{p}|$ for all $\Lambda\rightarrow \pi^+ \mu^-$ MC]{(Color Online) $\Delta tof$ \textrm{vs} $p$ for $\Lambda\rightarrow \pi^+ \mu^-$ Monte Carlo.  The red curves define the boundaries of the PID cuts for each particle type.  Shown are the recoil $K^+$ (left), decay $\pi^+$ (center), and decay $\mu^-$ (right).  The bands with $\Delta tof> 1.0$~ns in the kaon plot represent $\mu^+$ from kaon decays.  %The shading axis has a logarithmic scale.
}
\label{f:kpem_dtof}
\end{center}
\end{figure*}

\subsection{Example: $\Lambda \rightarrow \pi^{+} \mu^-$}
Here, we demonstrate this process with the $\Lambda \rightarrow \pi^+ \mu^-$ channel; other channels, with the exception of the $K^{0}_{s}\nu$ decay, are analyzed using the same observables.

For the three charged tracks in each event, we must first decide which positive tracks correspond to the recoil $K^{+}$ and $\pi^+$.
We make this assignment by calculating the invariant mass of each positive-negative track pair, and choosing the assignment that gives a value nearest the nominal $\Lambda$ mass.
We then apply PID cuts based on the $\Delta tof$ method to each of the three particles, the boundaries of which are shown in Fig.~\ref{f:kpem_dtof}.
These cuts are loose, and we have found that their efficiencies for each particle type are similar when applied to broader MC and data samples.

After the PID cuts have been made, we turn to analysis cuts based on kinematic observables.
For the $\Lambda \rightarrow \pi^{+}\mu^{-}$ channel, we make a symmetric cut on $MM^2$ (centered about $0$ with width $w_{1}$) and $MM(K^{+}_{r})$ (centered about 1.1186~GeV$/c^2$, with width $w_{2}$).
In order to find the widths, $w_{1}$ and $w_{2}$, which optimize the Punzi metric, we uniformly sample forty values for each width, resulting in 1600 distinct pairs with
\begin{eqnarray}
w_1 &\in& [0,0.001]\textrm{ GeV}^2/c^4\\
w_2 &\in& [0,0.03]\textrm{ GeV}/c^2.
\end{eqnarray}
We apply these cuts to both signal MC and data, and inspect the resulting $INV(\pi^{+},\mu^-)$ histograms with the signal region of the data histogram blinded.
We define the signal region to be values of $INV(\pi^{+},\mu^-)$ within 0.03~GeV$/c^2$ of the nominal $\Lambda$ mass (\textit{i.e.}, [1.086,1.146]~GeV/$c^2$), and the side-band regions to be within 0.15~GeV$/c^2$ of the peak (excluding the signal region).

\begin{figure*}
\begin{center}
\includegraphics[width=0.98\textwidth]{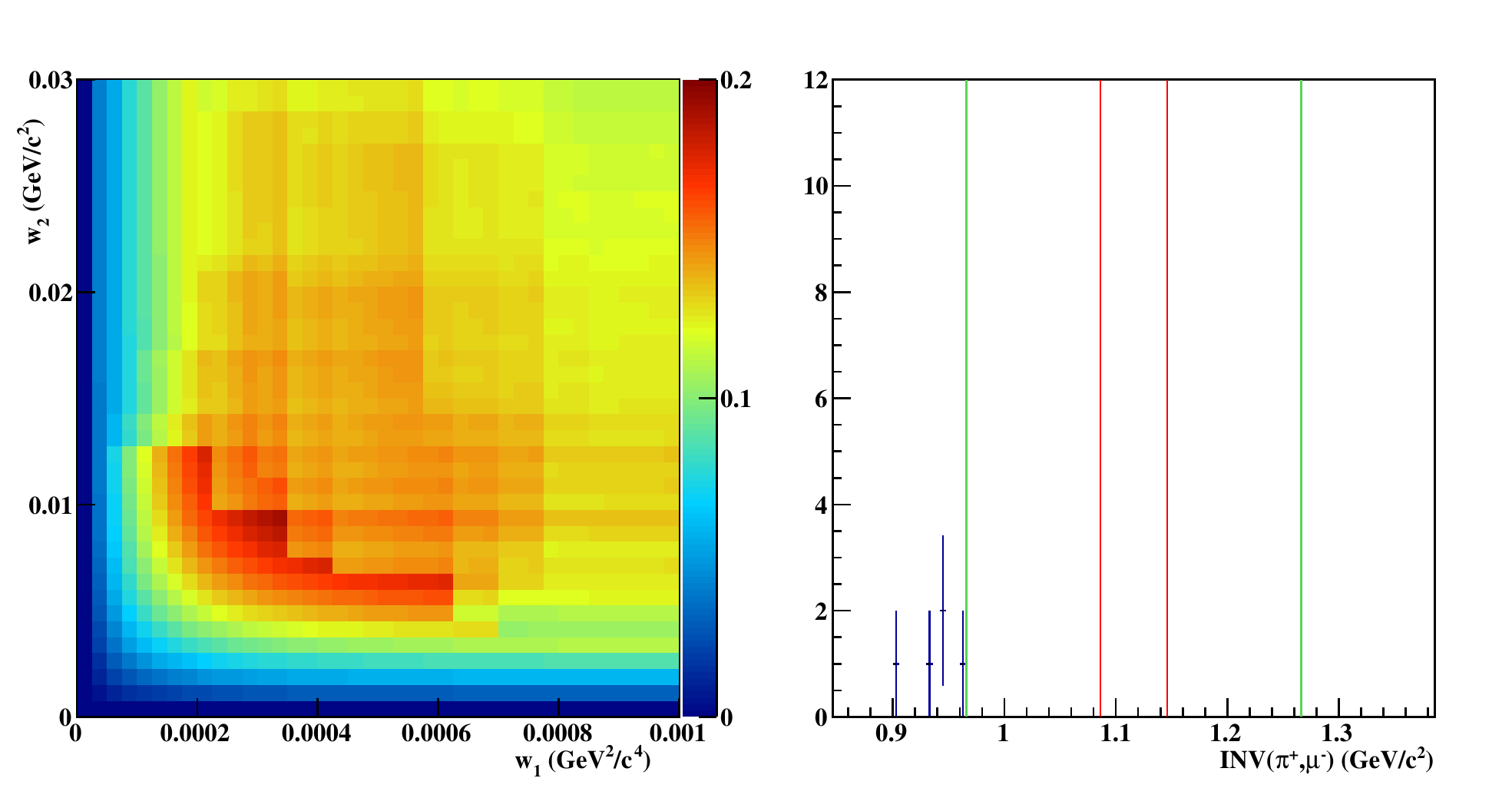}\\
\vspace{-0.4cm} (a) \hspace{0.45\textwidth} (b)\\
\vspace{0.1cm}\includegraphics[width=0.98\textwidth]{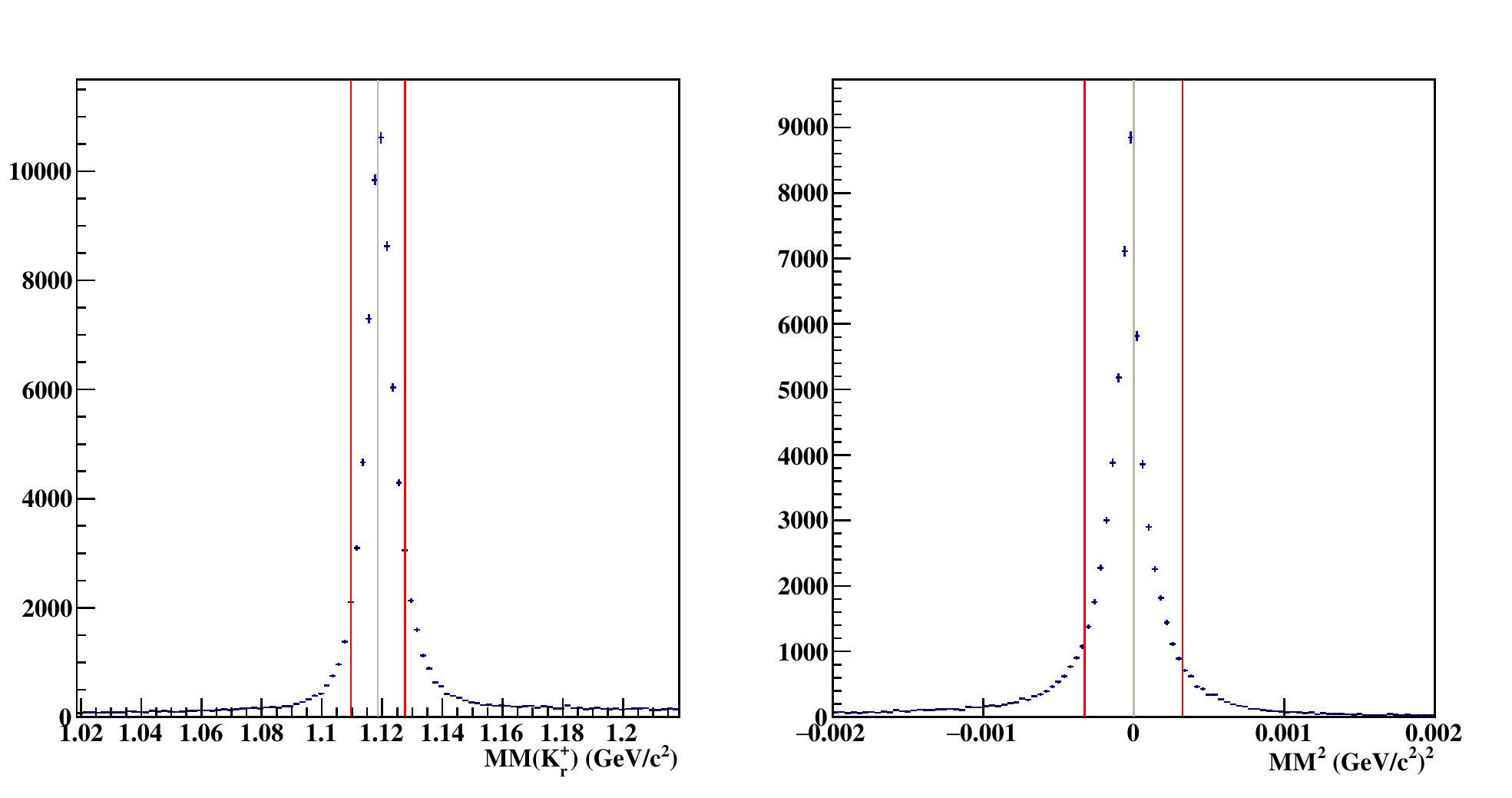}\\
\vspace{-0.4cm} (c) \hspace{0.45\textwidth} (d)\\
\caption[$\mathcal{P}$ as a function of $w_1$ (GeV$^2/c^4$) and $w_2$ (GeV$/c^2$) for the $\Lambda\rightarrow K^{+} e^-$ BNV channel]{(Color Online) Plots of $\mathcal{P}$ {\em vs} cut widths $w_1$ (GeV$^2/c^4$) and $w_2$ (GeV$/c^2$) (a), data $INV(\pi^+ \mu^-)$ (b), Monte Carlo $MM(K^+)$ (c), and Monte Carlo $MM^2$ (d) for the $\Lambda\rightarrow \pi^+ \mu^-$ channel. 
In (b), the vertical red lines show the boundaries of the blinded signal region, and the green vertical lines show the boundaries of the side-band regions.
In (c) and (d), the grey and red lines show the center and boundaries of the optimal cuts, respectively.}
\label{f:pvws}
\end{center}
\end{figure*}

We then use the signal MC distributions before and after cuts to determine the signal efficiency, $\epsilon(C)$.
We apply a simple side-band technique to the data $INV(\pi^{+},\mu^-)$ histogram to extrapolate $B(C)$, the expected number of background events in the blinded signal region.
We then use these values to calculate $\mathcal{P}$ for each width pair (see Fig.~\ref{f:pvws}), and select the width pair that maximizes $\mathcal{P}$ as optimal.
%Properties of optimal cuts are shown in Table~\ref{t:cutprops}.
For the $\Lambda \rightarrow \pi^+ \mu^-$ channel, we find the optimal widths to be $w_1 = 3.25\times10^{-4}$~GeV$^2/c^4$ and $w_2 = 9.00\times10^{-3}$~GeV$/c^2$.
The plots in Fig.~\ref{f:pvws} illustrate the resulting cuts and the blinded signal plot for this channel.

Fig.~\ref{f:pvws} demonstrates that the optimal cut scheme for this channel is quite restrictive; only five data events populate the (blinded) $INV(\pi^+,\mu^-)$ histogram, and none of these falls within the side-band regions.
We thus estimate the expected number of background events in the signal region to be zero.
By studying the signal MC for this channel, we estimate that the efficiency of these cuts, $\epsilon(C)$, is 7.91\% (including the effects of detector acceptance).
The effects of detector acceptance and analysis cuts are similar for all of the charged decay channels.
Table~\ref{t:ul0} lists the properties of the optimal cuts for each channel.

%% \begin{table}
%% \caption[Properties of optimal BNV selection cuts]{Properties of optimal BNV selection cuts.  $w_1$ and $w_2$ denote the widths of cuts on $MM^2$ and $MM(K^{+}_{r})$.  $\epsilon$ denotes the signal efficiency of these cuts based on application to signal MC.  $B$ denotes the number of background events in the signal region expected from sideband studies.}\label{t:cutprops}
%% \begin{center}
%% \begin{tabular}{lcccc}\hline
%% \multirow{2}{*}{Reaction} & $w_1$ & $w_2$ & $\epsilon$ & \multirow{2}{*}{$B(C)$} \\
%%  & $(\textrm{GeV}/c^2)^2$ & (GeV$/c^2$) & ($\times10^{-2}$) \\ \hline \hline
%% $\Lambda \rightarrow K^{+} e^-$ & $2.50\times10^{-4}$ & 0.01625 & 4.13 & 0\\
%% $\Lambda \rightarrow K^{+} \mu^-$ & $3.25\times10^{-4}$ & 0.0125 & 4.42 & 0\\

%% $\Lambda \rightarrow K^{-} e^+$ & $1.80\times10^{-3}$ & 0.01375 & 4.63 & 0\\
%% $\Lambda \rightarrow K^{-} \mu^+$ & $3.00\times10^{-4}$ & 0.0300 & 4.40 & 0\\

%% $\Lambda \rightarrow \pi^{+} e^-$ & $2.75\times10^{-4}$ & 0.00900 & 7.02 & 0\\ 
%% $\Lambda \rightarrow \pi^{+} \mu^-$ & $3.25\times10^{-4}$ & 0.00900 & 7.91 & 0\\
%% $\Lambda \rightarrow \pi^{-} e^+$ & $4.75\times10^{-4}$ & 0.0125 & 8.65  & 0.75\\
%% $\Lambda \rightarrow \pi^{-} \mu^+$ & $3.50\times10^{-4}$ & 0.00900 & 7.92 & 0.25\\

%% $\Lambda \rightarrow \bar{p} \pi^+$ & $5.00\times 10^{-4}$ & 0.0425 & 4.98 & 0\\
%% \end{tabular}
%% \end{center}
%% \end{table}

\subsection{$\Lambda \rightarrow \bar{p} \pi^+$}
%% The $\Lambda \rightarrow \bar{p}\pi^+$ reaction presents an opportunity to search for $\Lambda$-$\bar{\Lambda}$ osciallations, {\em i.e.} a process by which the $\Lambda$ oscillates into its anti-particle counterpart ($\bar{\Lambda}$) which then undergoes a standard-model decay ($\bar{\Lambda}\rightarrow \bar{p}\pi^+$).
%% Though there is not a simple way to picture this reaction proceeding via $X$ boson coupling, theoretical and experimental ({\em e.g.}~\cite{Serebrov2008181,Abe:2011ky}) work has been performed by other groups looking for similar oscialltions of the neutron.  
%% Such baryon-antibaryon oscillations would have far-reaching implications and are often held up as evidence for high-energy theories ranging from see-saw models~\cite{babu,dutta} to extra dimensions~\cite{nussinov}. 

Our analysis of the $\Lambda\rightarrow \bar{p}\pi^+$ channel proceeds nearly identically to that of the charged $\Lambda \rightarrow m \ell$ channels.
We first assign the positively charged track mass hypothesis by comparing $INV(\bar{p}, \pi^+)$ for the two possible track permutations.
We then apply a PID cut in the $\Delta tof$ {\em vs} momentum plane for each particle.
Threshold $\bar{p}$ photoproduction via the process $\gamma p \rightarrow p p \bar{p}$ occurs at a photon energy of 3.751~GeV; the maximum tagged photon energy in our dataset is 3.86~GeV.
Unlike in the other charged decay channels where there are significant numbers of each particle type present in the data, the null hypothesis suggests that there should be relatively few anti-protons in the dataset.
As a result, we use a less restrictive PID cut for the anti-proton, keeping events with $\Delta tof$ between -1.8~ns and 1.0~ns.
Because of the absence of background reactions that produce $\bar{p}$, this PID cut is the most stringent requirement in this channel.  
We optimize cuts on the $MM^2$ and $MM(K^+)$ observables in the same method as for the $\Lambda \rightarrow m \ell$ channels, yielding optimal cut widths of $w_1 = 5.00\times10^{-4}$~GeV$^2/c^4$ and $w_2 = 4.25\times 10^{-2}$~GeV$/c^2$, respectively, and a signal efficiency of $\epsilon = 4.98\%$.

\subsection{$\Lambda \rightarrow K^{0}_{S}\nu$}
In addition to the charged decay modes, we search for the decay of $\Lambda \rightarrow K^{0}_{S}\nu$, using the dominant charged decay mode $K^{0}_{S}\rightarrow \pi^+\pi^-$.
(Observation of the $K^0\rightarrow 2\pi$ selects $K^{0}_{S}$ rather than $K^{0}_{L}$.)
Because of this reaction's final-state neutrino, we do not have access to $INV(K^{0}_{S},\nu)$; thus, the analysis described for the charged decay modes is not appropriate for this channel.
In addition, the unmeasured momentum of the $\nu$ limits our analysis constraints and we can expect more background to pass the optimized cuts.

We begin the background separation process by applying two-dimensional PID cuts based on $\Delta tof$ and momentum to the charged final state particles (recoil $K^+$ and the $\pi^\pm$ from $K^0$ decay) similar to those for the $\Lambda \rightarrow \pi^+\mu^-$ channel (see Fig.~\ref{f:kpem_dtof}).
We then optimize a two-dimensional cut motivated by the particulars of this decay.
The first is a symmetric cut on $MM^2$, centered at 0 (the mass of the $\nu$ is negligible) with width $w_1$.

\begin{figure*}
\begin{center}
\includegraphics[width=0.98\textwidth]{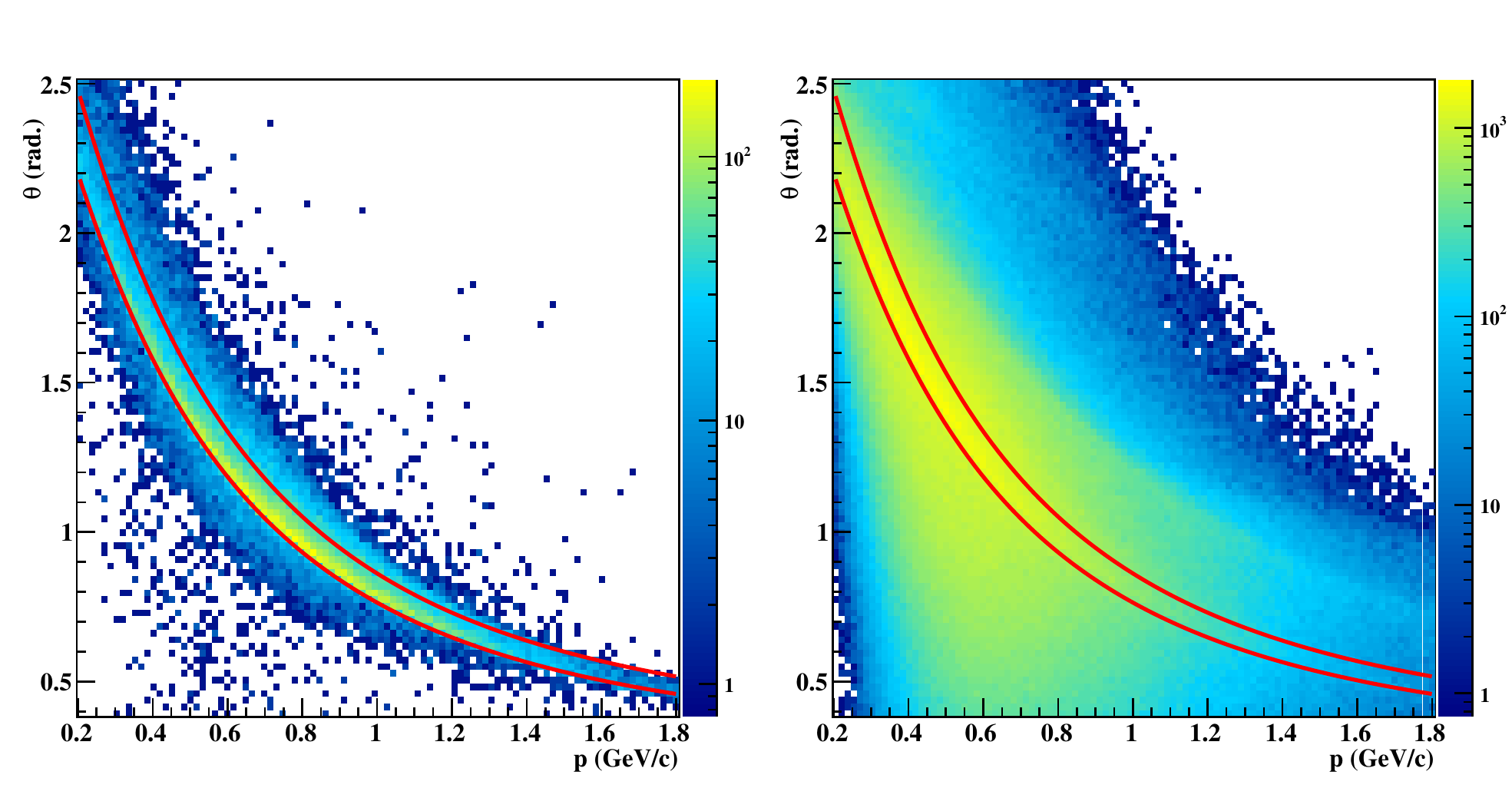}\\
\caption[$\theta_{o}^{\pi}$ vs $p$]{(Color Online) $K^{0}_{S}\rightarrow \pi^+\pi^-$ opening angle {\em vs} $p$ for $\Lambda\rightarrow K^{0}_{S} \nu$ Monte Carlo (left) and data (right) after PID cuts.  
The red curves demonstrate the boundaries of the optimized cut.}
\label{f:ks_open}
\end{center}
\end{figure*}

The second cut identifies $\pi^\pm$ pairs that are produced from $K^0$ decay by inspecting the $\pi^{\pm}$ opening angle, ($\theta^{\pi}_{o}$), in the c.m.~frame:
\begin{equation}
\theta^{\pi}_{o} = \cos^{-1}\left(\frac{\vec{p}_{+}\cdot\vec{p}_{-}}{|\vec{p}_{+}||\vec{p}_{-}|} \right),
\end{equation}
where $\vec{p}_\pm$ are the momenta of the decay pions indexed by charge.
Due to the break-up energy associated with the $K^0\rightarrow \pi^+ \pi^-$ decay, $\theta^{\pi}_{o}$ is constrained to a narrow range for a given value of $K^0$ momentum.
However, for $\pi^+\pi^-$ pairs that do not come from $K^0$ decay, we expect only the constraints associated with momentum conservation for the entire event, {\em i.e.} much less correlation between the pion momenta.
Distributions of $\theta^{\pi}_{o}$ {\em vs} magnitude of $K^0$ momentum, $p_{K}$, for data and signal MC are shown in Fig.\ref{f:ks_open}.
We separate $K^0\rightarrow\pi^+\pi^-$ events by making a two-dimensional cut on $\theta^{\pi}_{o}$ \textit{vs}~$p_{K}$.
To obtain a description of the correlation between $\theta_{o}^{\pi}$ and $p_{K}$, we fit the two-dimensional histogram of the two observables and found adequate description with the function
\begin{equation}
f(p_{K}) = 0.319 - \frac{0.459}{p_{K}+0.5} + \frac{2.23}{(p_{K}+0.5)^{2}} - \frac{0.641}{(p_{K}+0.5)^{3}}.
\end{equation}
The cut width, $w_{2}$, is implemented by keeping events for which
\begin{equation}
(1-w_2)f(p_{K}) \leq \theta^{\pi}_{o} \leq (1+w_2)f(p_{K}).
\end{equation}

\begin{figure}
\begin{center}
\includegraphics[width=0.49\textwidth]{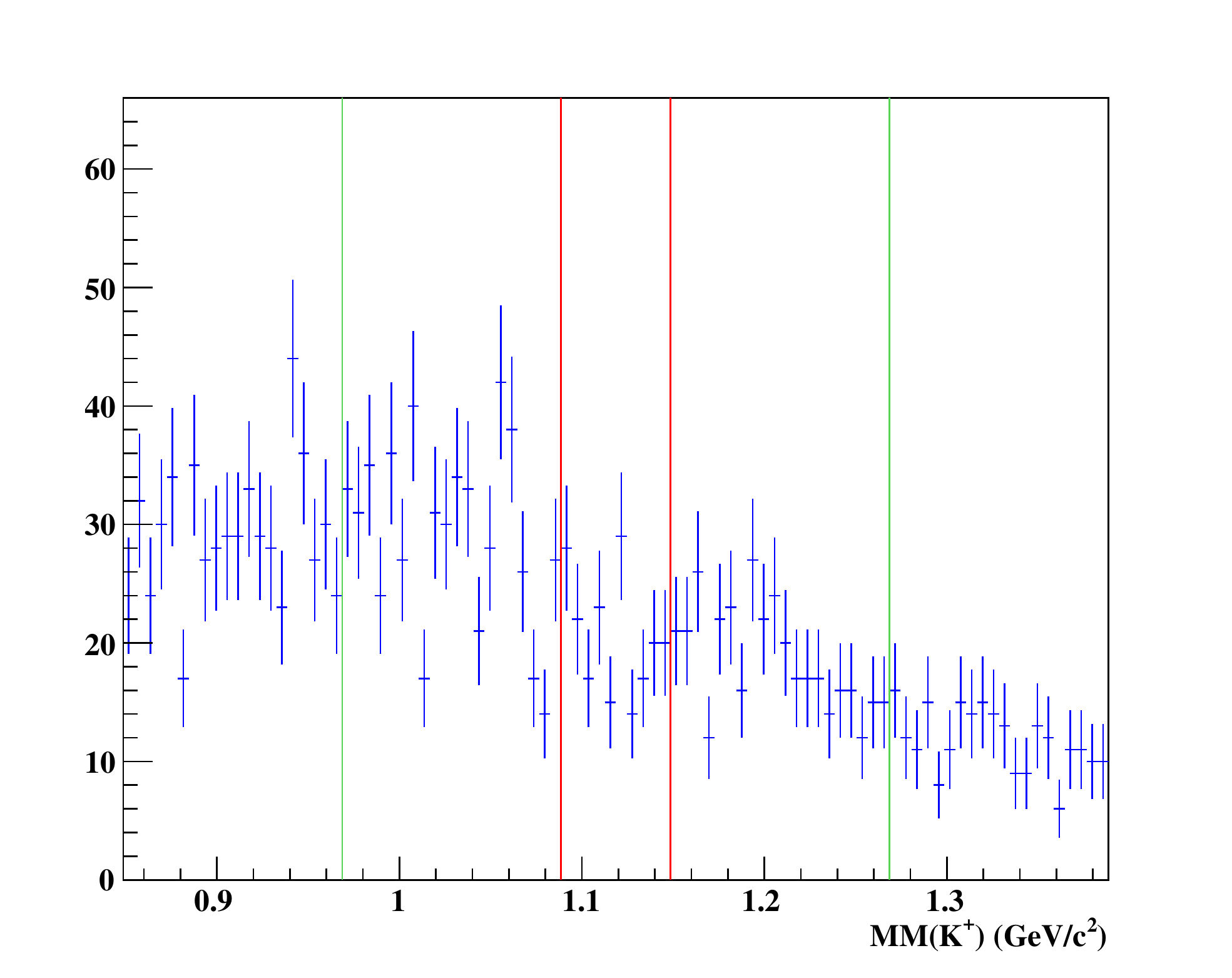}\\
\caption[Unblinded $MM(K^+)$ for $\Lambda \rightarrow K^{0}_{S}\nu$ decay]{(Color Online) Unblinded $MM(K^+)$ distribution for data events after application of cuts optimized for the $\Lambda\rightarrow K^{0}_{S} \nu$ channel.
The red vertical lines show the boundaries of the blinded signal region, and the green vertical lines show the boundaries of the side-band regions.}
\label{f:mmoffk_ksnu}
\end{center}
\end{figure}

The optimization process tests 1600 pairings of cut widths with $w_1 \in [0,0.15]\textrm{ GeV}^{2}/c^{4}$ and $w_2 \in [0,0.2]$.
Because $INV(K^0, \nu)$ is not accessible, we use the $MM(K^+)$ distributions to estimate $\epsilon(C)$ and $B(C)$, and ultimately to identify the signal.
We choose the blinded signal region to be within $0.03$~GeV$/c^2$ of the $\Lambda$ mass peak in the $MM(K^+)$ spectrum determined from standard model data and MC, 1.186~GeV$/c^2$, and the side-band regions to be within 0.15~GeV$/c^2$ of the peak (excluding the signal region).
We find the optimal widths to be $w_1 = 1.875\times 10^{-2}$~GeV$^{2}/c^{4}$ and $w_2 = 0.0600$, yielding an efficiency of 2.23\% and 239.25 estimated background events in the signal region.
The $MM(K^+)$ distribution for data events after unblinding is shown in Fig.~\ref{f:mmoffk_ksnu}.

%% \begin{table}
%% \caption[Properties of optimal BNV selection cuts]{Properties of optimal BNV selection cuts.  $w_1$ and $w_2$ denote the widths of cuts on $MM^2$ and $MM(K^{+}_{r}$.  $\epsilon$ denotes the signal efficiency of these cuts based on application to signal MC.  $B$ denotes the number of background events in the signal region expected from sideband studies.}\label{t:cutprops}
%% \begin{tabular}{lcccc}\hline
%% \multirow{2}{*}{Reaction} & $w_1$ & $w_2$ & $\epsilon$ & \multirow{2}{*}{$B$} \\
%%  & $(\textrm{GeV}/c^2)^2$ & (GeV$/c^2$) & ($\times10^{-2}$) \\ \hline \hline
%% $\Lambda \rightarrow K^{+} e^-$ & $7.175\times10^{-4}$ & 0.101575 & 5.13 & 0.0 \\
%% $\Lambda \rightarrow K^{+} \mu^-$ & $1.05\times10^{-3}$ & 0.02335 & 5.20 & 0.0 \\
%% $\Lambda \rightarrow K^{-} e^+$ & $2.49\times10^{-3}$ & 0.033175 & 5.14 & 0.0 \\
%% $\Lambda \rightarrow K^{-} \mu^+$ & $5.35\times10^{-4}$ & 0.028225 & 4.94 & 0.25 \\
%% $\Lambda \rightarrow \pi^{+} e^-$ & $8.35\times10^{-4}$ & 0.017225 & 9.55 & 0.0 \\ 
%% $\Lambda \rightarrow \pi^{+} \mu^-$ & $6.95\times10^{-4}$ & 0.0364 & 10.5 & 0.5 \\
%% $\Lambda \rightarrow \pi^{-} e^+$ & $5.90\times10^{-4}$ & 0.0187 & 9.23 & 0.0 \\
%% $\Lambda \rightarrow \pi^{-} \mu^+$ & $3.10\times10^{-4}$ & 0.04525 & 8.96 & 0.0 \\
%% $\Lambda \rightarrow \bar{p} \pi^+$ & 0.0288925 & 0.017625 & 5.09 & 0.0 \\
%% \end{tabular}
%% \end{table}

\subsection{Assessing background signatures}

Because of the sensitive nature of a positive signal identification, understanding the signature (shape) of the background in signal identification histograms is crucial.
If a peak is present when the data in unblinded, we must be sure that the peak represents BNV signal, and is not merely an unfortunate distortion of the background events due to the cuts used.
In order to assess the signature of the background events in $INV(A,B)$ and $MM(K^{+})$ distributions, we generated $5\times10^5$ Monte Carlo events for each of six background reactions:
\begin{itemize*}
\item $\gamma p \rightarrow K^+\Lambda \rightarrow K^+ p \pi^-$
\item $\gamma p \rightarrow K^+\Sigma^0 \rightarrow K^+ p \pi^- \gamma$
\item $\gamma p \rightarrow p e^+ e^-$
\item $\gamma p \rightarrow p \mu^+ \mu^-$
\item $\gamma p \rightarrow p \pi^+ \pi^-$
\item $\gamma p \rightarrow p K^+ K^-$
\item $\gamma p \rightarrow p \pi^+ \pi^- \pi^0$
\end{itemize*}
These reactions were chosen either for their abundance in the dataset (combination of large cross section and detectability in CLAS) or for their similarity to the BNV channels investigated (similar final-state particles).
After applying the optimized cuts for each BNV channel to these MC events, we found that very few events pass the cuts; no channel's cuts allow more than one background MC event into the signal region of the $INV(A,B)$ or $MM(K^+)$ histograms.
Thus, we claim that none of the background reactions investigated create an excess in the signal regions.

%\section{\boldmath Extraction of results\label{sec:extraction}}
\section{\boldmath Results\label{sec:results}}
%\input{extraction_of_results.tex}
%%%%%%%%%%%%%%%%%%%%%%%%%%%%%%%%%%%%%%%%%%%%%%%%%%%%%%%%%%%%%%%%%%%%%%%%%%%%%%%%
% Interpretation of results
%%%%%%%%%%%%%%%%%%%%%%%%%%%%%%%%%%%%%%%%%%%%%%%%%%%%%%%%%%%%%%%%%%%%%%%%%%%%%%%%
\subsection{Charged decays}

After the selection criteria are finalized using the Monte Carlo and side-band
studies, we applied these cuts to the unblinded data. For the nine decay modes
where the final state can be completely reconstructed, we found the number of observed events in the signal region, $N_{\textrm{obs}}$, to be between 0 and 2, consistent with background estimates from the cut optimization studies. 
For these decay modes we used the Feldman-Cousins approach~\cite{feldcous}
to determine upper limits on the reconstructed signal yields, $N_{\textrm{UL}}$.
%We then calculate an upper limit on the branching fraction for each BNV channel as
%\begin{equation}
%\mathcal{B}_{\textrm{UL}} = \mathcal{B}_{p\pi}\frac{N_{\textrm{UL}}}{\epsilon N_{\textrm{prod}}},
%\end{equation}
%where $\mathcal{B}_{p\pi}$ is the branching fraction for the charged standard model decay, $\Lambda \rightarrow p\pi^-$; $\epsilon$ is the signal efficiency of the optimized cuts; and $N_{\textrm{prod}}$ is taken from eq.~\ref{e:nprod}.

The Feldman-Cousins approach provides a way to estimate upper confidence
limits for null results. The inputs are the expected number of background
events ($N_{\rm eb}$) and the observed number of events ($N_{\rm obs}$).
We estimated the expected number of background events from the side-bands in 
the data (see previous section). 
%We note that the sideband estimates are consistent with Monte
%Carlo studies, and suggest that we would expect to find 1 or 0 background events in the signal region. 
$N_{\textrm{obs}}$ and $N_{\textrm{eb}}$ values for each decay mode are shown in Table~\ref{t:ul0}.
Fractional numbers are given when only one event was observed in a side-band
region that spans a greater range than the signal region. 
%In one decay mode ($\pi^{+} \mu^-$)
%we observe 1 event and 0 for all others, consistent with background expectations.

These provided the input to the Feldman-Cousins method and we quote the upper limit
on the reconstructed signal yield at 90\% confidence level ($N_{\rm UL}$), shown in Table~\ref{t:ul0}.
To calculate the upper limit on the branching fraction ($\mathcal{B}_{\rm UL}$),
we used the efficiency ($\epsilon$) for each decay mode as determined from Monte Carlo studies and the
total number of $\gamma p\rightarrow K^+\Lambda\rightarrow K^+ p \pi^-$ events produced during the data-taking period, $N_{\textrm{prod}}$ (see Eqn.~\ref{e:nprod}):
\begin{eqnarray}
    \mathcal{B}_{\rm UL} &=& \mathcal{B}_{p \pi} \frac{N_{\rm UL}}{\epsilon N_{\textrm{prod}}},
\end{eqnarray}
where $\mathcal{B}_{p\pi} = 0.639\pm0.005$ is the $\Lambda\rightarrow p \pi^-$ branching fraction~\cite{PDG-2014}.
The $\mathcal{B}_{\textrm{UL}}$ values for all charged decay modes are shown in Table~\ref{t:ul0}.

\begin{table*}
\caption[Events observed and upper limits]{Inputs to and final calculations for the 
upper limits on the branching fractions.
$w_1$ and $w_2$ are the optimized cut widths for each channel.  
For the charged decay modes, $w_1$ and $w_2$ are the widths of the cuts on $MM^2$ (in GeV$^2/c^4$) and $MM(K^+)$ (in GeV$/c^2$), respectively.
For $\Lambda\rightarrow K^0_{S}\nu$, $w_1$ and $w_2$ are the widths of the cuts on $MM^2$ (in GeV$^2/c^4$) and $K_{S}^{0}\rightarrow \pi^+\pi^-$ opening angle, respectively.
$\epsilon$ is the efficiency for each reaction, calculated using Monte Carlo studies described 
in the text, $N_{\textrm{eb}}$ is the number of expected background events, calculated
from side-band studies and confirmed with Monte Carlo studies, $N_{\rm obs}$ is the number
of observed events in the data, $N_{\rm UL}$ is an upper limit on the number of signal events
in the data, calculated using the Feldman-Cousins technique~\cite{feldcous} for the charged decay modes and likelihood scanning technique for the $\Lambda \rightarrow K^{0}_{S}\nu$ channel, and $\mathcal{B}_{\rm UL}$
is the upper limit on the $\Lambda$ branching fraction for each decay mode.}
\label{t:ul0}
\begin{tabular}{lccccccc}
    %\multirow{2}{*}{Reaction} & $\epsilon$ (\%) & $N_{\textrm{eb}}$ & $N_{\rm obs}$ & $N_{\rm UL}$ & $\mathcal{B}_{\rm UL} (\times 10^{-7}$) \\
\hline
Reaction & $w_1$ & $w_2$ & ~\ $\epsilon$ (\%) ~\ & ~\ $N_{\textrm{eb}}$ ~\ & ~\ $N_{\rm obs}$ ~\ & ~\ $N_{\rm UL}$ ~\ & $\mathcal{B}_{\rm UL}$ \\
\hline
\hline
$\Lambda \rightarrow K^{+} e^-$ & $2.50\times10^{-4}$ & 0.01625 &       4.13 &  0 &     1 & 4.36 & $2\times10^{-6}$ \\
$\Lambda \rightarrow K^{+} \mu^-$ & $3.25\times10^{-4}$ & 0.0125 &      4.42 &  0 &     2 & 5.91 & $3\times10^{-6}$ \\
$\Lambda \rightarrow K^{-} e^+$ & $1.80\times10^{-3}$ & 0.01375 &       4.63 &  0 &     1 & 4.36 & $2\times10^{-6}$ \\
$\Lambda \rightarrow K^{-} \mu^+$ & $3.00\times10^{-4} $ & 0.0300 &     4.40 &  0 &     2 & 5.91 & $3\times10^{-6}$ \\
$\Lambda \rightarrow \pi^{+} e^-$ & $2.75\times10^{-4}$ & 0.00900 &     7.02 &  0 &     0 & 2.44 & $6\times10^{-7}$\\
$\Lambda \rightarrow \pi^{+} \mu^-$ & $3.25\times10^{-4}$ & 0.00900 &   7.91 &  0 &     0 & 2.44 & $6\times10^{-7}$ \\
$\Lambda \rightarrow \pi^{-} e^+$ & $4.75\times10^{-4}$ & 0.0125 &      8.65 &  0.75 &  0 & 1.94 & $4\times10^{-7}$ \\
$\Lambda \rightarrow \pi^{-} \mu^+$ & $3.50\times10^{-4}$ & 0.00900 &   7.92 &  0.25 &  0 & 2.44 & $6\times10^{-7}$ \\
$\Lambda \rightarrow \bar{p}\pi^+$ & $5.00\times10^{-4}$ & 0.0425 &     4.98 &  0 &     0 & 2.44 & $9\times10^{-7}$ \\
$\Lambda \rightarrow K^0_S \nu$ & 0.01875 & 0.0600 & 2.23 & 239.25 & -3.88 & 14.1 & $2\times10^{-5}$ \\
\end{tabular}
\end{table*}

\subsection{$\Lambda \rightarrow K^{0}_{S}\nu$}
For the $K^0_S \nu$ decay mode, a BNV signal would manifest itself as a peak in 
the $MM(K^+)$ distribution at the $\Lambda$ mass. 
When we unblinded the $MM(K^+)$ histogram, we observed no such peak and found a number of events in the signal region that is consistent with the background study above.
The number of events in the signal region is much larger than is normally handled by the Feldman-Cousins approach; thus, we perform a likelihood scan to determine the upper limit on $N_{\textrm{UL}}$.

We performed an unbinned maximum likelihood fit to the data in
this region using an exponential function probability density function (PDF)
to describe the background and a sum of two Gaussians to describe the signal.
The shape of the background was allowed to vary in the fit, as are the 
numbers of signal and background events. 
The parameters describing the Gaussians ({\em i.e.} means and widths)
were fixed to values determined from Monte Carlo studies.

The fit converged to a central value of $N_{\rm obs}=-3.88\pm 8.9$ signal events,
consistent with 0 signal events. To check
whether or not this negative value is of concern, we sampled from 
a distribution described by the background parameters returned by the fit
to generate 1000 mock ``background-only" samples
and fit them to a background-plus-signal hypothesis. About 50\% of these fits
returned a negative value for the signal and about 35\% returned a value 
more negative than what was found in the data. We determined that the negative 
value is an artifact of fitting to a small number of points using a function 
with as much freedom as we use. We note that nowhere does the total PDF
go negative. 

To calculate an upper limit on the signal yield, we scaned the likelihood
function by performing a series of fits where
the signal yield ($N$) is varied around the best fit value
$N_{\rm obs}$ and the other parameters were refit to map out
the difference in the $\ln$-likelihood: $\Delta \ln \mathcal{L} =  \ln
\mathcal{L}(N_{\rm obs})-\ln \mathcal{L}(N)$. We
integrated the function $y = e^{-\Delta \ln \mathcal{L}}$ over $N$.
We ignored the unphysical region with $N<0$ and calculate the
integral for $N>0$. We note the value of $N$ which 
encloses 90\% of the area above $N=0$ 
and interpret this as the upper limit on the signal yield returned by the fit
at 90\% confidence level. This procedure returns an upper limit ($N_{\textrm UL}$) of 14.1
signal events.

\subsection{Experimental uncertainties}

Uncertainty in $\mathcal{B}_{\textrm{UL}}$ comes from the world average of the $\Lambda\rightarrow p \pi^+$
branching fraction ($0.8\%$)~\cite{PDG-2014} and statistical and systematic uncertainties from the extraction of $N_{\textrm{prod}}$ and cut efficiencies for each BNV channel.
We found a 6.1\% relative uncertainty in $N_{\textrm{prod}}$ by combining the 0.02\% systematic uncertainty due to the $INV(p,\pi^-)$ peak fitting procedure, the $\approx 6\%$ systematic uncertainty in CLAS acceptance calculation (taken from previous hyperon production analysis~\cite{mccracken}), and the 0.64\% statistical uncertainty in $\epsilon_{p\pi^-}$ (estimated with binomial statistics).
We estimated the uncertainty in $\epsilon$ for each BNV channel by comparing the effects of optimized $MM(K^+)$ and $MM^2$ cuts on MC and standard-model data distributions, and found it to be $\approx 7.6\%$.
We combined all uncertainties in quadrature to find a relative uncertainty in $\mathcal{B}_{\textrm{UL}}$ of $\approx 9.8\%$.
With this estimate of the combined uncertainty in hand, we quote the final $\mathcal{B}_{\textrm{UL}}$ results to one significant figure (see Table~\ref{t:ul0}).

\section{Summary\label{sec:summary}}
%%%%%%%%%%%%%%%%%%%%%%%%%%%%%%%%%%%%%%%%%%%%%%%%%%%%%%%%%%%%%%%%%%%%%%%%%%%%%%%%
% Conclusion
%%%%%%%%%%%%%%%%%%%%%%%%%%%%%%%%%%%%%%%%%%%%%%%%%%%%%%%%%%%%%%%%%%%%%%%%%%%%%%%%
The analysis described here represents the first search for baryon- and lepton-number violating decays of the $\Lambda$ hyperon.
Though similar studies have been performed with much higher sensitivities for decays of the nucleon, 
this study offers the first direct probe of BNV processes involving strange quarks in the initial state.
Using a dataset for photoproduction off of the proton collected with the CLAS detector at Jefferson Laboratory containing roughly $1.8\times 10^{6}$ reconstructed Standard Model $\Lambda \rightarrow p \pi^-$ decays, we have searched via blinded analysis for BNV decays of the $\Lambda$ to either meson-lepton pairs or to $\bar{p}\pi^+$.
We found no BNV signal in any of the ten decay channels investigated, and set upper limits on branching fraction for each of the processes studied in the range $7\times 10^{-7}$ to $2\times 10^{-5}$.

%%%%%%%%%%%%%%%%%%%%%%%%%%%%%%%%%%%%%%%%%%%%%%%%%%%%%%%%%%%%%%%%%%%%%%%%%%%%%%%%
%%%%%%%%%%%%%%%%%%%%%%%%%%%%%%%%%%%%%%%%%%%%%%%%%%%%%%%%%%%%%%%%%%%%%%%%%%%%%%%%
% Input the pubboard acknowledgements file
\section{Acknowledgements}
We are grateful for the excellent luminosity and machine conditions provided by the staff and administration of the Thomas Jefferson National Accelerator Facility.
This work was supported in part by the U.S. Department of Energy (under grant No. DE-FG02-87ER40315); the National Science Foundation; the Italian Istituto Nazionale di Fisica Nucleare; the French Centre National de la Recherche Scientifique; the French Commissariat \`{a} l'Energie Atomique; an Emmy Noether Grant from the Deutsche Forschungsgemeinschaft; the U.K. Research Council, S.T.F.C.; and the National Research Foundation of Korea.  
The Southeastern Universities Research Association (SURA) operated Jefferson Lab under United States DOE contract DE-AC05-84ER40150 during this work.

% or this one for PRDs
%\input pubboard/acknowledgements.tex
\vfill
%%%%%%%%%%%%%%%%%%%%%%%%%%%%%%%%%%%%%%%%%%%%%%%%%%%%%%%%%%%%%%%%%%%%%%%%%%%%%%%%
%%%%%%%%%%%%%%%%%%%%%%%%%%%%%%%%%%%%%%%%%%%%%%%%%%%%%%%%%%%%%%%%%%%%%%%%%%%%%%%%
%\bibliographystyle{unsrt}
\bibliographystyle{apsrev4-1}
\bibliography{note}

\end{document}